\documentclass[prd,aps,amsmath,nofootinbib,amssymb,eqsecnum,showkeys,tightenlines,12pt]{revtex4-1}

\usepackage{amsmath, latexsym, amssymb, graphicx, color, multirow, bm, braket, ascmac, bbm}
\usepackage{romannum}
\usepackage{hyperref}
\usepackage[caption=false]{subfig}
\usepackage{dsfont}
\usepackage[table]{xcolor}
\usepackage{tabularx}
\usepackage[T1]{fontenc}
\usepackage{mathtools}
\usepackage[capitalize]{cleveref}
\usepackage{comment}
\usepackage{ulem}
\definecolor{nicered}{rgb}{.7,.1,.1}
\definecolor{nicegreen}{rgb}{.1,.5,.1}
\definecolor{cobalt}{rgb}{0,0,.5}
\definecolor{cornellred}{rgb}{0.7,0.11,.11}
\definecolor{Alizarin}{rgb}{0.82,0.1,.26}
\definecolor{seagreen}{rgb}{0.7, 0.75, 0.71}
\definecolor{darkblue}{rgb}{0.0, 0.0, 0.55}
\hypersetup{colorlinks,citecolor=nicegreen,linkcolor=nicered, urlcolor=darkblue}
\usepackage{multirow}
\usepackage{slashed}
\numberwithin{equation}{section}
\usepackage{verbatim}
\usepackage{autobreak}

\newcommand{\hl}[1]{{\color{cyan} #1}}

\begin{document}

\title{Probing Electroweak Phase Transition in the Singlet Standard Model via $bb\gamma\gamma$ and 4$l$ channels
	\\
}
\author{Wenxing Zhang$^{1}$, Hao-Lin Li$^{2}$, Kun Liu$^{1}$, Michael J. Ramsey-Musolf$^{1,3}$, Yonghao Zeng$^{1}$, Suntharan Arunasalam$^{1}$}
\affiliation{\vspace{2mm} \\
	$^1$Tsung-Dao Lee Institute and School of Physics and Astronomy, \\
            Shanghai Jiao Tong University, 800 Dongchuan Road, Shanghai 200240, China \\
 	$^2$Centre for Cosmology, Particle Physics and Phenomenology (CP3),  Universite Catholique de Louvain, \\
	$^3$Amherst Center for Fundamental Interactions, Department of Physics, University of Massachusetts, Amherst, MA 01003, USA
}

\begin{abstract}
	We investigate the prospects for resonant di-Higgs and heavy Higgs production searches at the 14 TeV HL-LHC in the combination of $bb\gamma\gamma$ and $4l$ channels, as a probe of  a possible first order electroweak phase transition in real singlet scalar extension of the Standard Model. 
		Event selection follows those utilized in the $bb\gamma\gamma$ and $4l$ searches by the ATLAS Collaboration, applied to simulation using benchmark parameters that realize a strong first order electroweak phase transition.
		The output of discriminant analysis is implemented  by numerical calculation, optimised by the  joint restriction from the two channels.
		The prospective reach for $bb\gamma\gamma$/$4l$ channel could be more competitive in probing the electroweak phase transition at lower/higher resonance masses. 
		With 3 $ab^{-1}$ integrated luminosity, the combination of the $bb\gamma\gamma$ and $4l$ channels can discover/exclude a significant portion of
		of the viable parameter space that realizes a strong first order phase transition when the resonance mass is heavier than 500 GeV.

\end{abstract}
\maketitle

\section{Introduction}
\pagenumbering{arabic}

As the last piece of the Standard Model(SM), the Higgs boson was discovered in 2012 at the Large Hardron Collider(LHC)~\cite{ATLAS:2012yve, CMS:2012qbp}. 
The SM has achieved great success in describing a plethora of electroweak and strong interaction phenomena. However, the SM fails to explain a number of observable facts about the universe. Among these is the origin of the cosmic baryon asymmetry $Y_B=n_b/s$, whose present value is $Y_B = \left(8.59 \pm 0.11\right) \times 10^{-11}$ according to the PLANCK measurement~\cite{Planck:2013pxb}. This open puzzle provides one of the strong motivations to chase after a theory of physics beyond the SM.

One of the most well-studied and experimentally testable baryogenesis scenarios is electroweak baryogenesis (EWBG)\cite{Kuzmin:1985mm, Shaposhnikov:1986jp, Shaposhnikov:1987tw, Cohen:1993nk}, wherein $Y_B$ is generated in conjunction with electroweak symmetry breaking (EWSB). For a review and references, see Ref.~\cite{Morrissey:2012db}.
As with other baryogenesis scenarios, successful EWBG
must satisfy the three "Sakharov criteria"~\cite{Sakharov:1967dj, Kajantie:1996qd}: 
(1) baryon number violation; (2) C and CP-violation; and (3) departure from thermal equilibrium. 
The first two conditions guarantee that the net baryon number generation is possible. 
The last one is to ensure the net baryon number not be "washed out" by sphaleron process~\cite{Klinkhamer:1984di}. In EWBG, (3) is satisfied by the occurrence of a 
strong first order electroweak phase transition (SFOEWPT), which provides the conditions under which new CP-violating interactions generate the asymmetry.
Within the SM, given the known mass of Higgs boson, the phase transition is a smooth crossover~\cite{Kajantie:1996qd, Kajantie:1996mn}. 
To explain the magnitude of $Y_B$, new physics, especially those including new scalars coupling to the SM Higgs field, is essential to lead to a strong SFOEWPT. Based on general considerations, the presence of a first order EWPT implies that these particles must be lighter than $\sim 800-1000$ GeV and couple with sufficient strength to the Higgs boson, making them and their interactions a clear target for high energy collider studies\cite{Ramsey-Musolf:2019lsf}.


The simplest realization of a SFOEWPT is achieved through adding one real singlet scalar in Higgs sector~\cite{OConnell:2006rsp,Profumo:2007wc,Kotwal:2016tex,Hashino:2016xoj}, which is called the real singlet-extended standard model (xSM). 
The scenario admits two pathways of EWSB: (i) a direct transition to the electroweak broken vacuum (\lq\lq Higgs phase\rq\rq), 
and (ii) a two-step first-order transition with the singlet scalar acquiring a vacuum expectation value (vev) before electroweak symmetry breaking. In either case, the singlet scalar may also obtain a vev in the Higgs phase.
In the presence of a non-vanishing singlet vev, the singlet-like mass eigenstate and the SM-like mass eigenstate mix with each other after EWSB. 
Therefore, the LHC experiments searching for heavy scalar resonance shed light on exploring the viable parameter space for EWPT in the xSM~\cite{Ramsey-Musolf:2019lsf}. 
Such experiments can be classified into several categories:
(\romannum{1}) Di-Higgs channels with final states including 4$b$~\cite{ATLAS:2018rnh, CMS:2018qmt}, $bbVV^*$~\cite{ATLAS:2018fpd, CMS:2017rpp, ATLAS:2017tlw}, $bbl\nu \bar{l}\bar{\nu}$~\cite{CMS:2017rpp}, $WW^* WW^*$~\cite{ATLAS:2018ili}, bb$\tau\tau$~\cite{ATLAS:2018uni, CMS:2017hea} and $bb\gamma\gamma$~\cite{ATLAS:2018dpp, CMS:2018tla}.
(\romannum{2}) Di-boson channels with semileptonic final states~\cite{ATLAS:2020fry, ATLAS:2017jag}, hadronic final states~\cite{ATLAS:2016hal, ATLAS:2017zuf} and leptonic final states~\cite{ATLAS:2018sbw, ATLAS:2017uhp, ATLAS:2017tlw}.
(\romannum{3}) Di-fermion channels~\cite{ATLAS:2020zms}.
On the other hand, studies have been performed for the parameter space that can realise the SFOEWPT.
Heavy resonance mass up to 500 GeV can be observed by $bb\gamma\gamma$ and 4$\tau$ searches at the 14 TeV HL-LHC with lumiosity equal to 3 ab$^{-1}$~\cite{Kotwal:2016tex}.
Also, in search for signals with $bbWW^*$ final state, the singlet-like scalar mass can reach 350 GeV at the 13 TeV, 3 ab$^{-1}$ LHC~\cite{Huang:2017jws}.
For recent study in xSM, see in Ref.~\cite{Espinosa:2011ax, Patel:2011th, Laine:1994zq, Ekstedt:2018ftj, Garny:2012cg, Carena:2019une, Kozaczuk:2019pet, Alves:2018jsw, Kurup:2017dzf, Chen:2017qcz, Kotwal:2016tex, Chen:2014ask, Katz:2014bha, Damgaard:2015con, Huang:2015tdv, Curtin:2014jma, No:2013wsa, Huang:2012wn, Damgaard:2013kva, Ashoorioon:2009nf, Blasi:2022woz, Ghorbani:2020xqv, Ghorbani:2018yfr}.

Importantly, the singlet-like scalar will decay to conventional SM-like Higgs decay products due to singlet-doublet mixing as well as to di-Higgs final states (if sufficiently heavy). Thus, it is natural to ask how the combination of such \lq\lq heavy Higgs\rq\rq\, searches and resonant di-Higgs searches can further probe the SFOEWPT-viability in the xSM. 
In this paper, we focus on the HL-LHC collider experimental tests on the xSM, including the $pp \to h_2 \to h_1 h_1 \to bb\gamma\gamma$ and $pp \to h_2 \to Z Z \to 4\ell$ final states. 
Our strategy in doing so is as follows. 
We perform a parameter scan by requiring parameters to satisfy the theoretical bounds, including perturbativity and stability of the potential, and fix the mass and vev of SM Higgs doublet to the observed values.
From the scanned parameter points, we further restrict the doublet-singlet mixing angle $\theta$ by considering the current global Higgs searches and constraints from electroweak precision observables (EWPO)\footnote{A recent measurement of the $W$ boson mass by the CDF-II collaboration deviates significantly the previous world average, which leads to difficulties in combining the measurement with previous results. Thus, in this work we will use the constraints from the fit of EWPO without including the newly measured $W$ mass.}.
We select benchmark points that produce maximum and minimum $bb\gamma\gamma$ cross-section with resonance mass ranging from 260 GeV to 800 GeV and can simultaneously produce the SFOEWPT in the early universe. 
To analyze the perspective sensitivity of HL-LHC on testing these benchmark points, we perform collider simulations of the two aforementioned processes by following ATLAS analysis~\cite{ATLAS:2018dpp, ATLAS:2020fry} and obtain the signal efficiencies and distributions, 
while The background distributions are simply rescaled from the result in the experimental paper~\cite{ATLAS:2018dpp, ATLAS:2020fry}.

As a preview, we summarise the main result in this paper:

\begin{itemize}
        \item  For resonance mass $\gtrsim 350$ GeV, the $4\ell$ channel demonstrates much stronger detection capabilities compared with the $bb\gamma\gamma$ channel.  For resonance mass $\gtrsim 600$ GeV, this channel covers the whole viable SFOEWPT parameter space in the xSM by $5\sigma$.
	\item  For resonance mass $\lesssim 350$ GeV, the $bb\gamma\gamma$ channel is relatively more powerful. The 95\% C.L. upper limit for the heavy resonance mass reaches 330 GeV and 480 GeV for parameter space with minimum and maximum $bb\gamma\gamma$ cross section respectively.
	\item  For the combined searches of the 4$\ell$ and $bb\gamma\gamma$ final states, if the resonance mass is located between 300 to 750 GeV, it is possible to observe or exclude the xSM SFOEWP-viable parameter space.  Moreover, the 95\% C.L. and $5\sigma$ upper limit lines cover the whole viable parameter space for the resonance mass $\gtrsim 550$ GeV and $\gtrsim$ 600 GeV respectively.   
        \item A strong correlation between the signs of the mixing angle $\theta$ and the heavy Higgs to SM-like Higgs coupling $g_{211}$ is observed for SFOEWPT-viable parameter space.
\end{itemize}

Our discussion of the analysis leading to these conclusions is organized as follows: In section~\ref{sec::model}, we introduce the xSM framework and the various restrictions on the parameter space: perturbativity, global Higgs searches and EWPO.
In section~\ref{sec::SFOEWPT}, we consider the one-loop effective potential with high-T approximation and perform a general scan over the allowed parameter space to obtain regions compatiable with a SFOEWPT. 
Section~\ref{subsec::constraints} performs a parameter scan with constraints from discribed above.
Benchmark paramater choices yielding minimum and maximum $bb\gamma\gamma$ cross sections and SFOEWPT are listed in the end of this section.
Section~\ref{sec::simulation} performs the $bb\gamma\gamma$ and 4$\ell$ simulation based on 13 TeV experiments and extrapolates the result to 14 TeV 3000fb$^{-1}$. 
Section~\ref{sec::analysis} presents the discrimination between signal and background events. Further it draws conclusions about the sensitivity of exclusion ability.
Section~\ref{sec::conclusion} is dedicated to the conclusions.
In the Appendix, subsection~\ref{sec::appdxrge} details the 1-loop RGE of xSM used for analyzing the perturbativity bound. Subsections~\ref{sec::appdx1} and \ref{sec::appdx2} demonstrate the details of the simulation of $bb\gamma\gamma$ and 4$\ell$ respectively. Subsection~\ref{sec::new_ewpo} calculates the impact from the current CDF-II W-boson mass measurement on the heavy Higgs mass and the mixing angle.

\section{The xSM model}\label{sec::model}

We consider a minimal extension of SM that includes a gauge singlet real scalar S.
The most general renormalizable scalar potential is given by~\cite{Profumo:2007wc}
\begin{align}\label{eq::TreeLevel}
V_0(H,S)&=-\mu^2 (H^\dagger H) + \lambda (H^\dagger H)^2 + \frac{a_1}{2} (H^\dagger H) S + \frac{a_2}{2} (H^\dagger H) S^2\\ \nonumber
&+ \frac{b_2}{2} S^2 + \frac{b_3}{3} S^3 + \frac{b_4}{4} S^4,
\end{align}
where $H$ is the SM $SU(2)_L$ scalar doublet. The fields $H$ and $S$ obtain vevs after spontaneous EWSB. They may be cast into the form

\begin{align}
	S&=x_0+s, \\
	H&=\left(
	\begin{matrix}
		G^+ \\ 
		\frac{1}{\sqrt{2}}(v_0+h+iG^0)
	\end{matrix}
	\right).
\end{align}

The $a_1$ and $a_2$ terms with a non-zero VEV of the singlet introduce mixing and a portal interaction between the SM Higgs and the singlet.  
The absence of $a_1$ and $b_3$ leads to a $Z_2$ symmetry that protects the singlet from mixing with SM particles if $x_0$ vanishes, and hence the singlet becomes a DM candidate.
However, even if the potential preserves the $Z_2$ symmetry explicitly, appropriate potential parameters can lead to a spontaneous $Z_2$ breaking, which yields a non-zero $x_0$ and mixes the SM Higgs and singlet scalar through the $a_2$ term, thereby allowing the decay of both $h_1$ and $h_2$ to SM particles.   For a recent analysis of this possibility, see Ref.~\cite{Carena:2019une}.
A first order EWPT may arise in several ways~\cite{Profumo:2007wc, Noble:2007kk, Espinosa:2007qk, Cline:2012hg, Profumo:2014opa} 
(a) a one-step transition to the present Higgs phase wherein $x_0=0$; 
(b) a one-step transition to a EWSB vacuum in which the vevs of both $H^0$ and $S$ are non-vanishing; 
(c) a two-step transition, where the vacuum in the first step has $H^0=0$ and $S\not=0$ while the vacuum at the end of the second step has $H^0\not=0$ and $S\not=0$. 
Case (a) requires thermal loop contributions from $S$ to generate a sufficiently large barrier between the symmetric and EWSB phases. 
For scenario (b), the cubic portal term can induce a tree-level barrier for $a_1<0$. 
The second step of the two-step history (c) will be first order due to the cross-quartic portal term with $a_2>0$. 
In the following, we study the SFOEWPT and its di-Higgs signal phenomenology.



The minimization (tadpole) conditions in xSM potential can be utilized to trade the two potential parameters $\mu^2$ and $b_2$ with vevs of the two scalars: 

\begin{align}\label{eq::tadpole}
\mu^2&=\lambda v_0^2+\left(a_1+a_2 x_0\right) \frac{x_0}{2},\\
b_2&=-b_3 x_0 - b_4 x_0^2 - \frac{a_1 v_0^2}{4x_0} - \frac{a_2 v_0^2}{2}.
\end{align}

The mass matrix of the scalar sector can be derived by taking the second derivatives of the scalar potential and evaluating their value at the scalar vevs:

\begin{equation}\label{eq:msMatrix}
\mathcal{M}^2=
\left(
\begin{matrix}
m_s^2 & m^2_{hs}\\
m^2_{hs} & m_h^2
\end{matrix}
\right).=
\left(
\begin{matrix}
-\frac{a_1 v_0^2}{4x_0}+x_0\left(b_3+2b_4 x_0\right) & \frac{v_0}{2} \left(a_1+2a_2 x_0\right)\\
\frac{v_0}{2} \left(a_1+2a_2 x_0\right) & 2 \lambda v_0^2
\end{matrix}
\right).
\end{equation}
The above scalar mass matrix can be diagonalized by the mixing matrix parameterized by the mixing angle $\theta$:
\begin{equation}
	O(\theta)^T \mathcal{M}^2 O(\theta) = \left(
	\begin{matrix}
		m_{1} & 0 \\
		0 & m_{h_2} 
	\end{matrix}
	\right),
	~~~~~O(\theta) = \left(
	\begin{matrix}
		\cos\theta & -\sin\theta \\
		\sin\theta & \cos\theta
	\end{matrix}
	\right).
\end{equation} 
where $m_{h_1}=125.25~\text{GeV}$~\cite{ParticleDataGroup:2022pth} corresponds to the mass of SM-like mass eigenstate. $h_2$ is the singlet-like mass eigenstate, such that the Higgs fields can be expressed as 
\begin{align}
	h_1&=h \cos\theta + s \sin\theta \\
	h_2&=s \cos\theta - h \sin\theta ,
\end{align}

To guarantee the stability of the potential, the coefficient of the quartic field term should be positive definite, where the quartic field term is expressed as
\begin{equation}
    V(h,s) \supset \left(
    \begin{matrix}
         h^2 & s^2
    \end{matrix}
    \right) 
    \left(
    \begin{matrix}
         \frac{\lambda}{4} & \frac{a_2}{8} \\
         \frac{a_2}{8} & \frac{b_4}{4}
    \end{matrix}
         \right)
    \left(
    \begin{matrix}
     h^2 \\ s^2     
    \end{matrix}
     \right).
\end{equation}
The stability condition requires $a_2 \geq -2\sqrt{\lambda b_4}$. 
Furthermore we need to guarantee that the local potential minimum at $(v_0, x_0)$ is the global one, which is checked numerically by evaluating the potential at all possible extreme points.

\section{Electroweak Phase Transition}\label{sec::SFOEWPT}
The character of EWPT is understood in terms of the finite-T effective potential. It is well-known that in the conventional treatment, $V^{T\neq 0}_\mathrm{eff}$, which is derived from the gauge-dependent 1PI effective action, suffers from gauge-dependence~\cite{Patel:2011th, Garny:2012cg, Ekstedt:2018ftj, Laine:1994zq}. 
However, in the high temperature expansion, wherein the potential is expanded in powers of the temperature, the leading field- and temperature-dependent term that arises at $\mathcal{O}(T^2)$ is gauge-independent when evaluating the potential at the tree-level minimum of the fields according to the $\hbar$-expansion prescription~\cite{Patel:2011th}. For the same reason, the 1-loop  zero-temperature Coleman-Weinberg potential, $V_{CW}$ is also gauge independent.
Therefore, in this paper, we employ the gauge-independent high-T approximated potential, which is composed of the tree-level potential, zero-temperature Coleman-Weinberg potential, $V_{CW}$, and the gauge-independent finite-temperature corrections expanded up to $\mathcal{O}(T^2)$.

We start by introducing the zero-temperature Coleman-Weinberg potential as follows:
\begin{equation}\label{eq::CW}
	V_{CW}=\sum_k \frac{(-1)^{2s_k}}{64\pi^2}  g_k ~[M^2_{k}]^2\left(\ln \frac{M_{k}^2}{\Lambda^2}+\frac{3}{2}\right),
\end{equation}
wherein the summation contains all fields that interact with the scalar fields $h$ and $s$.
The $s_k$ is the spin of the particle, and $M_k$ is the mass. The $\Lambda$ is the renormalization scale, which is fixed to be $v_0$.

The high-T approximation for bosonic fields and fermion fields are expressed according to~\cite{Profumo:2007wc}

\begin{align}\label{eq::high-T}
	V_\mathrm{High-T}^\mathrm{scalar}&=\frac{g_s T^4}{2\pi^2}\left(-\frac{\pi^4}{45}+\frac{\pi^2}{12}\frac{m_k^2}{T^2}-\frac{\pi}{6}\left(\frac{(m_k^2)^{3/2}}{T^3}\right)-\frac{m_k^4}{32T^4}~\text{log}\left(\frac{m_k^2}{c_B T^2}\right)\right)\\
	V_\mathrm{High-T}^\mathrm{fermion}&=-\frac{g_f T^4}{2\pi^2}\left(-\frac{7\pi^4}{360}-\frac{\pi^2}{24}\frac{m_k^2}{T^2}-\frac{m_k^4}{32T^4}~\text{log}\left(\frac{m_k^2}{c_F T^2}\right)\right)\\
\end{align}
respectively, where the $g_{s/f}$ is the number of degrees of freedom for scalars/fermions. 
We ignore the field-independent terms and keep the gauge-independent contributions up to the leading order of temperature.
The high-T potential approximation then gives

\begin{equation}
	V_\mathrm{High-T} \simeq \frac{T^2}{24}\left(3M_g^2+M_h^2+M_s^2+ 6M_W^2+3M_Z^2\right)+\frac{T^2}{48}\left(12M_t^2\right),
\end{equation}
where the field-dependent Goldstone, Higgs, and scalar masses are given by
\begin{align}
	M_g^2&=-\mu^2+\lambda h^2+\frac{s}{2} \left(a_1 + a_2 s\right)\\
	M_h^2&=M_g^2+2\lambda h^2\\
	M_s^2&=b_2+\frac{a_2}{2} h^2 + 2 b_3 s + 3 b_4 s^2 .
\end{align}
Note that after EWSB $M_g^2=0$, as can be seen from the tadpole conditions.

In the leading order of the high-T approximation, field-dependent thermal corrections appear in terms of quadratic in the scalar fields. By themselves, these thermal mass contributions do not generate a barrier between the origin and the global minima. Such a barrier can arise as follows, corresponding to the scenarios identified in Sec.~\ref{sec::model}: (a) For the one-step direct transition to a pure Higgs phase, thermal loops, generated by the cross-quartic interaction proportional to $a_2$, can enhance the SM contributions to the $T h^3$-induced barrier; (b) for the  one-step transition to the mixed vev-vacuum, the $Z_2$ breaking interactions proportional to $a_1$ and $b_3$ can generate a tree-level barrier, whose effect becomes active as $T$ decreases due to the thermal mass contributions; (c) in the two-step scenario, the cross-quartic interaction proportional to $a_2$ generates a tree-level barrier between the pure singlet and Higgs (or mixed singlet-Higgs) vacuum; the effect of this barrier again depends on $T$ due to the thermal mass contributions to the potential. In the present analysis, we retain only the gauge-invariant thermal mass contributions to the potential, so our parameter choices do not include the direct one-step transition (a). We defer an analysis of this case to future work.
 

In practice,  we compute phase transition-relevant quantities using the CosmoTransitions package~\cite{Wainwright:2011kj}. 
We obtain the critical temperature $T_c$ at which the broken and unbroken phases are degenerate and the corresponding Higgs vev $v_c$. Both quantities are readily gauge invariant in the leading order high-$T$ approximation.
The condition for a strong first-order phase transition is approximately given by
\begin{equation}
\label{eq:strength}
	\frac{v_c}{T_c} \gtrsim 1.
\end{equation}

\section{Constraints on Parameters and numerical results}\label{subsec::constraints}
In this section, we discuss the details of our parameter scan and all the relevant theoretical and experimental constraints.
In the xSM, the potential includes seven parameters: $a_1$, $a_2$, $b_3$, $b_4$, $\lambda$, $\mu^2$, and $b_2$. By utilizing the tadpole condition described in Eqs.~\eqref{eq::tadpole}, it is possible to trade two parameters, $\mu^2$ and $b_2$, for the vacuum expectation values (vev) of scalars, denoted as $v_0$ and $x_0$. Additionally, we fix the SM Higgs vev $v_0$ to be 246 GeV and the mass of the SM-like Higgs $m_{h_1}$ to be 125.25 GeV~\cite{ParticleDataGroup:2022pth}, which reduces the number of free parameters in the theory to five.
Since the off-diagonal elements $m^2_{hs}$ in the mass matrix given by equation~\eqref{eq:msMatrix} have a simple linear relation to $a_2$, and $m^2_h$ and $m^2_s$ are independent of $a_2$, we can express $a_2$ in terms of the remaining parameters as
\begin{eqnarray}
\label{eq:a2other}
a_2 = \frac{\pm\sqrt{m_h^2m_s^2-m_{h_1}^2(m_h^2+m_s^2-m_{h_1}^2)}-a_1v_0/2}{x_0 v_0}.
\end{eqnarray}
In practice, we determine whether to choose the plus or minus sign in Eq.~(\ref{eq:a2other}) based on whether the resulting parameter point achieves the global minimum at $(v_0, x_0)$. If both signs satisfy the criteria, we randomly select one of the solutions. In the meantime, we also require the mass of the heavy Higgs $m_{h_2} > 2 m_{h_1}$ to enable the di-Higgs decay channel.

Consequently, we have five free parameters $a_1, b_3, b_4, \lambda, x_0$ to be scanned. We limit the range of the dimensionful parameters $a_1$, $b_3$, and $x_0$ to be within 1 TeV, specified as:
\begin{eqnarray}
-1 < \frac{a_1}{\text{TeV}}, \frac{b_3}{\text{TeV}} < 1, \quad 0 < x_0 < 1 ~\text{TeV}.
\end{eqnarray}
We choose this range due to the difficulty in finding parameter space that can yield a SFOEWPT with a heavy scalar boson mass $m_{h_2}$ exceeding 1 TeV\footnote{Note that the dearth of SFOEWPT-viable parameter points for $a_1$ and $b_3$ in this regime is consistent with the general arguments of Ref.~\cite{Ramsey-Musolf:2019lsf} }. As for $x_0$, we only scan the positive branch since the negative branch can be covered by reversing the signs of $a_1$ and $b_3$.

Regarding the dimensionless parameters $\lambda$ and $b_4$, 
we perform a test scan with their range to be within the naive parturbativity bounds --- $0 < \lambda, b_4 < 2\pi/3$~\cite{Lerner:2009xg}. 
As we have assumed the renormalization scale for these parameters are defined at the electroweak symmetry breaking scale $v_0$ in obtaining the zero-temperature Coleman-Weinberg potential,
we further check whether these dimensionless parameters after running to 10 TeV with one-loop RGEs\footnote{Running to higher scale results in stronger bounds.} are still within their naive perturbativity bounds: $0 < 6\lambda, 6b_4, |a_2| < 4\pi$. 
For those around 150,000 parameter points that survive from this perturbativity requirement after running, we find that the upper limits for the scanned dimensionless parameters $\lambda$ and $ b_4$ to be 1.052 and 0.868 respectively. 
Therefore to increase the scanning efficiency, we decided to specify ranges for dimensionless parameters to be narrower than given by the na\" ive perturbativity bounds:
\begin{eqnarray}
0 < \lambda, b_4 < 1.
\end{eqnarray}
We leave the details for full expressions of the renormalization group equations (RGEs) and the relevant values for input parameters in appendix.~\ref{sec::appdxrge}. 
During the parameter scan, we assume flat distributions for all the scanned parameters and trace the finite temperature potentials using CosmoTransitions for each obtained parameter point that satisfies the aforementioned theoretical constraints. Subsequently, we focus on selecting points that are capable of generating a  SFOEWPT in the early universe. Following this selection process, we find approximately 80,000 parameter points that meet the desired criteria out of 6 million scans.

As for the experimental bounds, the first constraint we consider is from the electroweak precision observables (EWPO). 
In the xSM, the oblique parameters receive additional contributions that depend on the singlet-like scalar mass $m_{h_2}$ and its mixing angle $\sin\theta$, and a constraint has been derived shows that the upper limit of $|\sin\theta |$ varies from 0.35 to 0.2 when the resonance mass satisfies $250~\text{GeV}<m_{h_2}<950~\text{GeV}$~\cite{Li:2019tfd}.
However, the new measurement of $m_W$ from CDF experiments indicates a very different parameter region from the previous measurement~\cite{ATLAS:2017rzl} and receives a lot of debate because of its deviation from the other measurements. 
Therefore, in what follows, we will utilize the EWPO constraints without including the new value of $m_W$ for reasons discussed in detail in Sec.~\ref{sec::new_ewpo}.

Secondly, we incorporate the constraints arising from the measurements of the signal strength of the Standard Model Higgs boson. The most recent results from ATLAS~\cite{ATLAS:2022vkf} and CMS~\cite{CMS:2020gsy} for LHC Run-2, assuming a global rescaling of the signal strength $\mu$, are as follows:
\begin{eqnarray}
\mu_{\rm ATLAS} = 1.05 \pm 0.06, \quad \mu_{\rm CMS} = 1.02^{+0.07}_{-0.06}.
\end{eqnarray}
By utilizing these measurements, we deduce that $|\sin \theta| < 0.193$ at a 95\% confidence level (CL). Thus, it becomes evident that the Higgs signal strength measurement imposes a more stringent constraint than the EWPO.

Apart from the SM-like Higgs, we also study the current and future prospective collider phenomenology for the heavy singlet-like resonance production at the 14 TeV HL-LHC.
Since $m_{h_2} > 2 m_{h_1}$, the decay of  $h_2 \to h_1 h_1$ and $h_2 \to X_{SM} X_{SM}$ is allowed,
where the $X_{SM}$ stands for SM particle whose mass is less than $m_{h_2}/2$.
Since the haevy Higgs $h_2$ interact with the SM particles through the mixing with the SM Higgs $h$, the total decay width of the heavy scalar $h_2$ can be written as
\begin{equation}
\Gamma_{h_2}=\Gamma_{h_2 \to h_1 h_1}+\sin^2\theta ~\Gamma^{\rm SM}(m_{h_2}),
\end{equation}
where the $\Gamma^{\rm SM}(m_{h_2})$ denotes the decay width function of standard model Higgs with a mass of $m_{h_2}$.
The decay width of $h_2 \to h_1 h_1$ in the equation above is
\begin{equation}
\Gamma_{h_2 \to h_1 h_1}=\frac{g_{211}^2 \sqrt{1-\frac{4m_{h_1}^2}{m_{h_2}^2}}}{8\pi m_{h_2}},
\end{equation}
where the tri-Higgs coupling is obtained from the tree-level potential:
\begin{align}\nonumber
g_{211}&=\frac{1}{4} [
\left(a_1+2a_2x_0\right)\cos^3\theta+4v_0\left(a_2-3\lambda\right)\cos^2\theta \sin \theta \\ \label{eq::g211}
&-2\left(a_1+2a_2x_0-2b_3-6b_4x_0\right)\cos\theta \sin^2 \theta - 2a_2 v_0 \sin^3 \theta ].
\end{align}

As a result, the branching ratio for $h_2 \to h_1 h_1$ can be expressed as  
\begin{equation}
	BR(h_2 \to h_1 h_1)=\frac{\Gamma_{h_2 \to h_1 h_1}}{\Gamma_{h_2 \to h_1 h_1}+\sin^2\theta ~\Gamma^{\rm SM}(m_{h_2})}.
\end{equation}

In this work, we focus on the di-Higgs channel with $bb\gamma\gamma$ final state and di-boson channel with four lepton final state.
The cross-section of $~p p \to h_2 \to h_1 h_1 \to bb\gamma\gamma$ and $~p p \to h_2 \to VV^* \to \ell\bar{\ell}\ell\bar{\ell}$ channels are calculated with narrow width approximation expressed in the following equations,
\begin{align}\label{eq::crosec}
\sigma_{bb\gamma\gamma}&=\sigma_{pp\to h_2}\times BR(h_2 \to h_1 h_1) \times BR(h_1 \to b \bar{b}) \times BR(h_1 \to \gamma \gamma), \\
\sigma_{4\ell}&=\sigma_{pp\to h_2}\times BR(h_2 \to ZZ) \times BR(ZZ \to 4\ell), 
\end{align}
where the production cross section for heavy Higgs can be expressed as $\sigma_{pp\to h_2}=\sin^2\theta \times \sigma_{H}(m_{h_2})$, and $\sigma_{H}(m_{h_2})$ is the SM Higgs production cross section if its mass were to be $m_{h_2}$ given by the CERN recommendation~\cite{Cepeda:2019klc}.

\begin{figure}[thb]
\centering
    \includegraphics[width=0.48\textwidth]{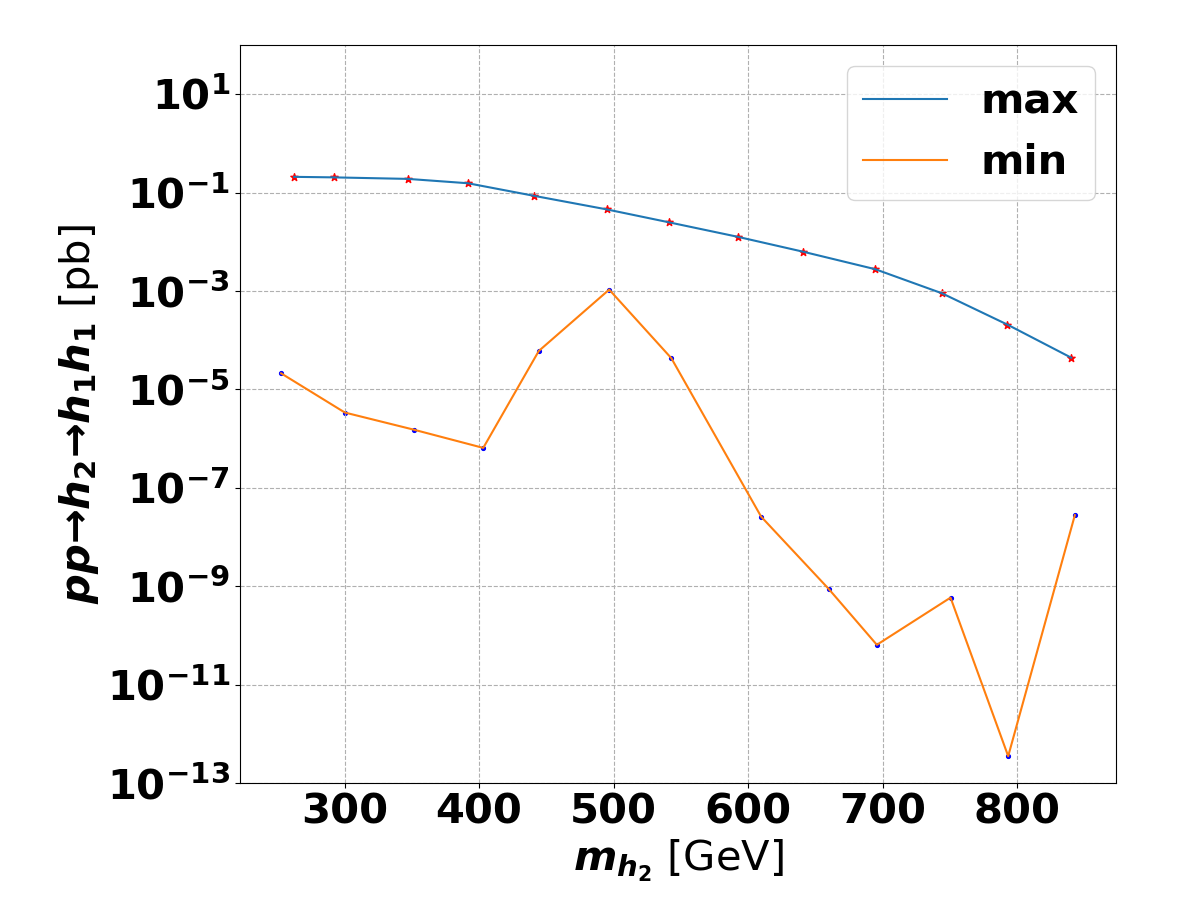}
    \includegraphics[width=0.48\textwidth]{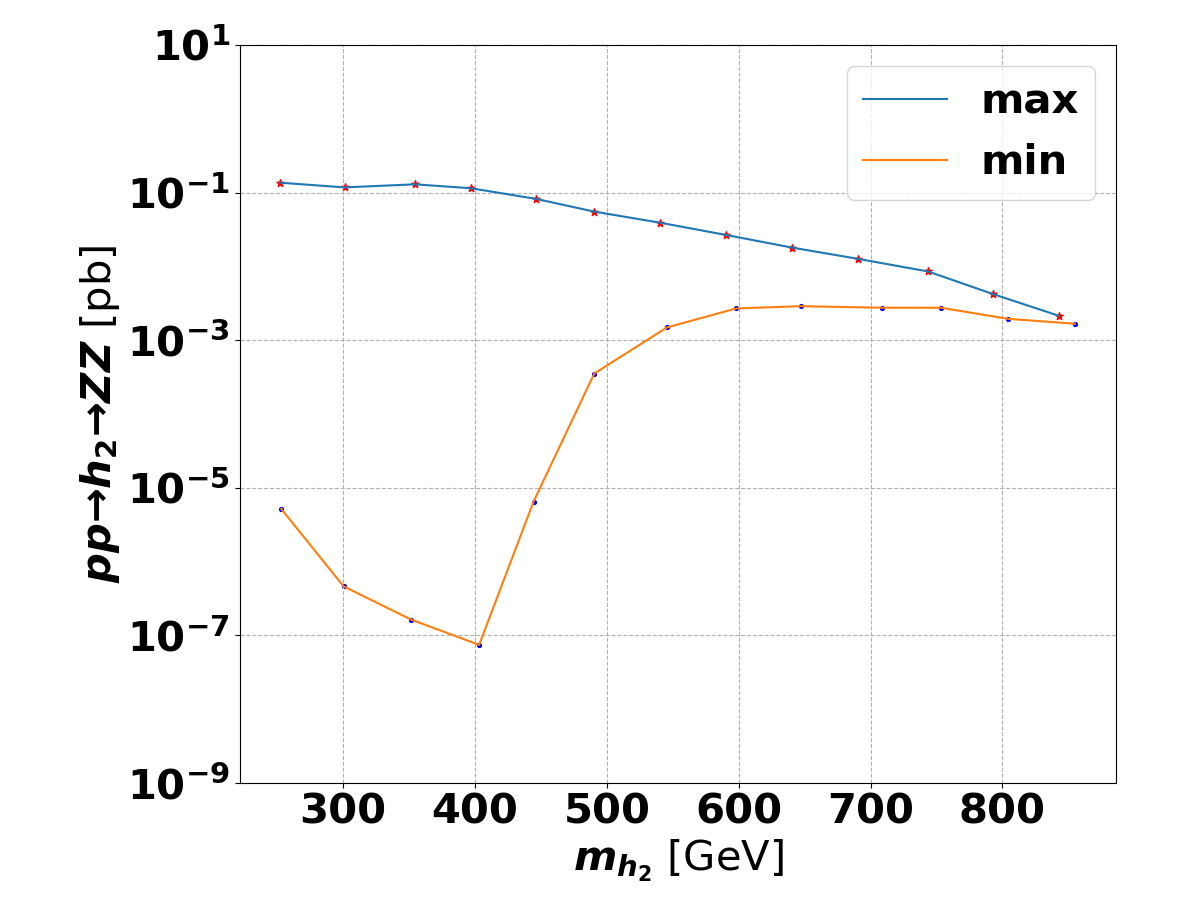}
    \includegraphics[width=0.48\textwidth]{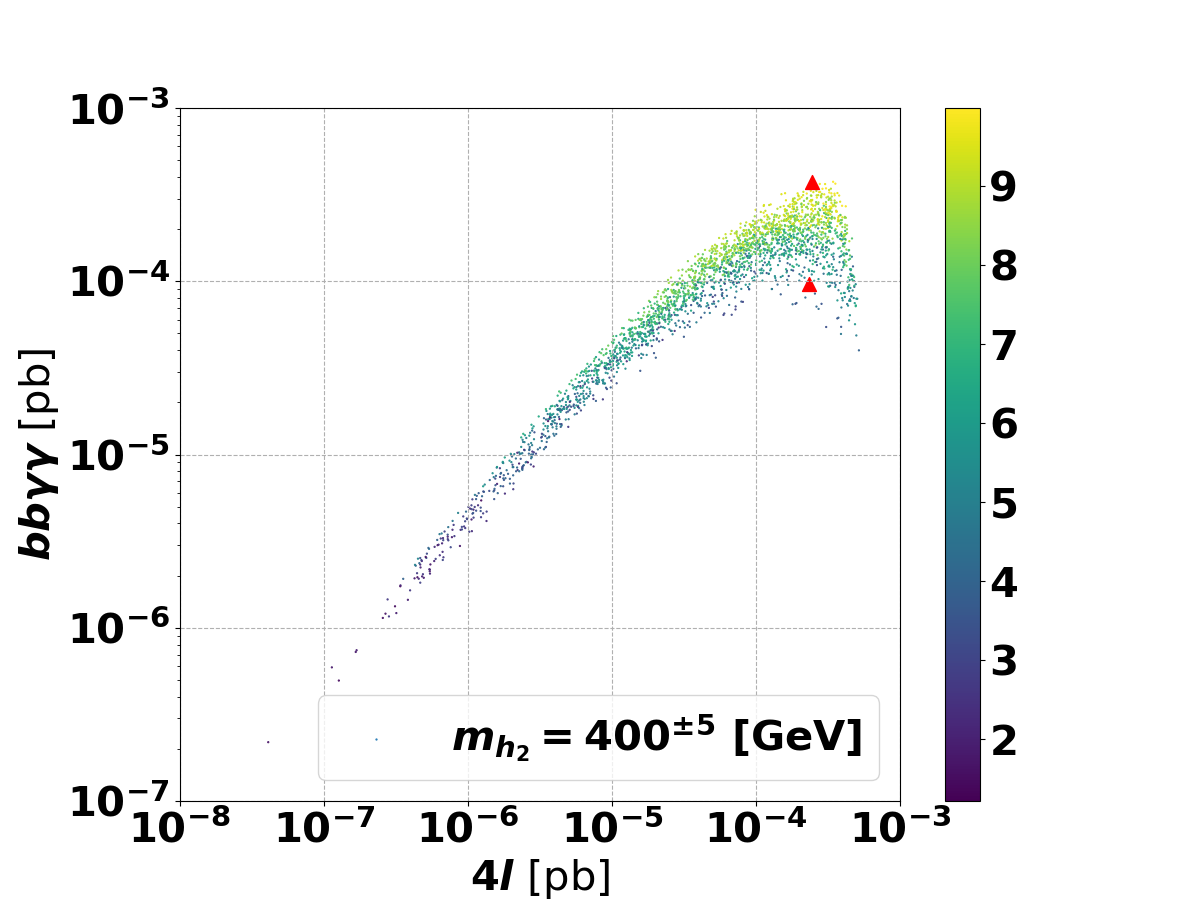}
    \caption{\label{fig::cro_sec}The maximum/minimum of the cross sections of $bb\gamma\gamma$ and 4l channels that realize SFOEWPT.  The color bar stands for first-order EWPT strength. The red points in the third panel are the maximum and minimum benchmarks we choose for $m_{h_2} = 402$ GeV and $m_{h_2} = 401$ GeV in the Tab.~\ref{tab:bchmk1} and Tab.~\ref{tab:bchmk2} respectively.}
\end{figure}

To estimate the ability of the HL-LHC in investigating EWPT in the xSM, we collect all the points in the scan that realize SFOEWPT. 
Among these points, for a given $h_2$ mass, the cross sections of bb$\gamma\gamma$ and 4$\ell$ channels vary from their minimum to the maximum depending on the choice of other parameters. In Fig.~\ref{fig::cro_sec}, we show the maximum and minimum cross sections for consecutive intervals of 50 GeV starting from 250 GeV.
Among selected points, benchmarks with given fixed 4$\ell$ cross sections that maximize/minimize bb$\gamma\gamma$ cross sections are listed in the table~\ref{tab:bchmk1}/table~\ref{tab:bchmk2}.

\begin{table*}[htbp]\centering
	\begin{tabular}{|c||c|c|c|c|c|c|c|c|c|c|c|} \hline
		Benchmark &$\sin\theta$ & $g_{211}$(GeV) & $m_{h_2}$(GeV) & $a_1$(GeV) & $a_2$ & $b_3$(GeV) & $b_4$ & $\lambda$ & $v_x$(GeV) &  $\sigma_{4l}~(pb)$ & $\sigma_{bb\gamma\gamma}~(pb)$  \\ \hline
		B1 & 0.181& 52& 262& -418& 2.64& -244& 0.647& 0.143& 78.5& 3.09$\times 10^{-4}$ & 5.50$\times 10^{-4}$\\ \hline
		B2 & 0.189& 60& 297& -435& 3.23& -365& 1.126& 0.162& 63.8& 2.89$\times 10^{-4}$ & 5.27$\times 10^{-4}$ \\ \hline
		B3 & 0.185& 74& 347& -554& 4.23& -361& 1.411& 0.210& 63.2& 3.01$\times 10^{-4}$ & 4.99$\times 10^{-4}$\\ \hline
		B4 &0.174& 91& 402& -664& 5.83& -470& 1.399& 0.291& 55.8& 2.43$\times 10^{-4}$ & 3.72$\times 10^{-4}$ \\ \hline
		B5 &0.167& 92& 445& -694& 6.29& -338& 1.093& 0.313& 50.5& 1.83$\times 10^{-4}$ & 2.13$\times 10^{-4}$\\ \hline
		B6 &0.152& 102& 495& -766& 8.16& -411& 1.569& 0.411& 45.1& 1.04$\times 10^{-4}$ & 1.159$\times 10^{-4}$\\ \hline
		B7 &0.152& 105& 546& -823& 9.51& -523& 1.856& 0.501& 40.0& 7.72$\times 10^{-5}$ & 6.08$\times 10^{-5}$ \\ \hline
		B8 &0.145& 106& 595& -876& 10.34& -34& 2.051& 0.561& 38.1& 5.06$\times 10^{-5}$ & 3.14$\times 10^{-5}$ \\ \hline
		B9 &0.137& 102& 646& -897& 11.25& 217& 1.995& 0.619& 33.7& 3.21$\times 10^{-5}$ & 1.49$\times 10^{-5}$\\ \hline
		B10 &0.150& 85& 696& -983& 10.91& 229& 2.039& 0.628& 31.9& 3.01$\times 10^{-5}$ & 6.02$\times 10^{-6}$ \\ \hline
  		B11 &0.142& 68& 748& -998& 11.32& 333& 1.974& 0.662& 28.0& 1.94$\times 10^{-5}$ & 2.06$\times 10^{-6}$ \\ \hline
	\end{tabular}
	\caption{\label{tab:bchmk1} Benchmark parameter choices that maximize the $bb\gamma\gamma$ channel cross section when the 4$\ell$ cross section is fixed. }
\end{table*}

\begin{table*}[htbp]\centering
	\begin{tabular}{|c||c|c|c|c|c|c|c|c|c|c|c|} \hline
		Benchmark &$\sin\theta$ & $g_{211}$(GeV) & $m_{h_2}$(GeV) & $a_1$(GeV) & $a_2$ & $b_3$(GeV) & $b_4$ & $\lambda$ & $v_x$(GeV) &  $\sigma_{4l}~(pb)$ & $\sigma_{bb\gamma\gamma}~(pb)$  \\ \hline
		B1 &0.135& 19& 262& -187& 1.52& 169& 0.766& 0.130& 50.5& 3.03$\times 10^{-4}$ & 5.64$\times 10^{-5}$\\ \hline
		B2 &0.155& 25& 299& -232& 1.78& 210& 0.847& 0.124& 46.7& 3.05$\times 10^{-4}$ & 1.41$\times 10^{-4}$\\ \hline
		B3 &0.146 & 33 & 353 & -291& 2.65& 174& 1.166& 0.165& 38.9& 2.90$\times 10^{-4}$& 1.36$\times 10^{-4}$\\ \hline
		B4 &0.142& 37& 403& -349& 3.11& 325& 0.748& 0.163& 36.0& 2.35$\times 10^{-4}$ & 9.64$\times 10^{-5}$\\ \hline
		B5 &0.142& 32& 447& -373& 3.39& 348& 1.229& 0.177& 30.6& 1.84$\times 10^{-4}$& 3.68$\times 10^{-5}$\\ \hline
		B6 &0.136& 42& 504& -467& 4.68& 265& 0.959& 0.235& 29.3& 1.05$\times 10^{-4}$ & 2.37$\times 10^{-5}$\\ \hline
		B7 &0.137& 40& 551& -528& 5.04& 544& 1.075& 0.272& 28.3& 7.66$\times 10^{-5}$ & 1.04 $\times 10^{-5}$\\ \hline
  	  B8 &0.135& 31& 603& -580& 5.44& 644& 1.525& 0.272& 25.8& 5.10$\times 10^{-5}$ & 3.17$\times 10^{-6}$\\ \hline
		B9 &0.129& 39& 651& -667& 7.24& 424& 1.926& 0.358& 24.8& 3.25$\times 10^{-5}$& 2.53$\times 10^{-6}$ \\ \hline
		B10 &0.149& 13& 704& -801& 6.28& 561& 1.384& 0.354& 25.8& 3.06$\times 10^{-5}$ & 1.39$\times 10^{-7}$\\ \hline
        B11 &0.141& 8& 754& -857& 6.94& 927& 1.655& 0.393& 24.2& 1.95$\times 10^{-5}$ & 3.09$\times 10^{-8}$ \\ \hline
	\end{tabular}
	\caption{\label{tab:bchmk2} Benchmark parameter choices that minimize the $bb\gamma\gamma$ channel cross section when the 4l cross section is fixed. }
\end{table*}

\begin{figure}[thb]
\centering
    \includegraphics[width=1.1\textwidth]{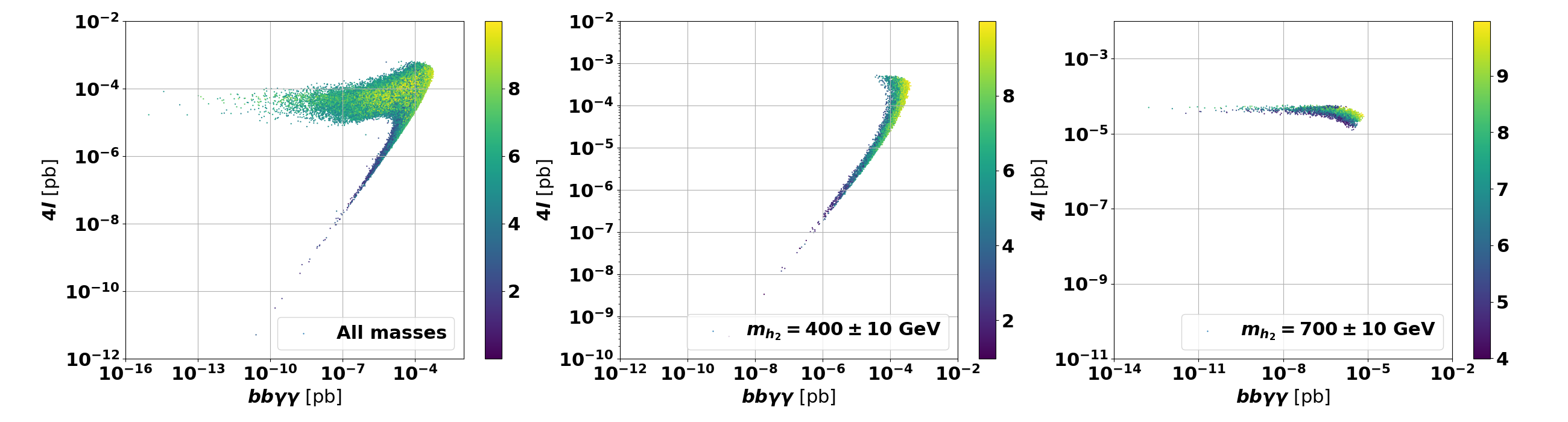}
	\caption{\label{fig::scan} Distribution of the cross sections for the $bb\gamma\gamma$ and 4$\ell$ channels. The color stands for first-order EWPT strength as defined in Eq.~(\ref{eq:strength}). The left panel shows all results with $m_{h_2}$ varying from 260 GeV to 800 GeV. The masses in the middle and right plots are within $400\pm 10$ GeV and $700\pm 10$ GeV, respectively. }
\end{figure}

The scanned result is illustrated in the Fig.~\ref{fig::scan}, where we plot the cross section for $p p \to h_2 \to h_1 h_1 \to b \bar{b} \gamma \gamma$ and $p p \to h_2 \to V V^{*} \to 4\ell$ to show their correlation.
In the following sections, we show the key results from our simulations for the two channels and derive the prospective bounds for future HL-LHC. 

\section{Simulation of signal and background distributions}\label{sec::simulation}
The strategy of signal and background simulation is as follows: two 13 TeV ATLAS analyses channels $p p \to h_2 \to h_1 h_1$~\cite{ATLAS:2018dpp} and $p p \to h_2 \to V V^{*}$~\cite{ATLAS:2017tlw} where $h_1 \to b \bar{b} /  \gamma \gamma$ and $V \to \ell \bar{\ell}$, are taken as references for event selection and background distribution. A factor of 1.18 is applied to backgrounds to count for collision energy upgrade from 13 TeV to 14 TeV, as commonly used in ATLAS prospective studies. The signal processes at 14 TeV center-of-mass energy are generated for heavy singlet-like mass states,  whose mass ranges from 270 GeV to 800 GeV in steps of 50 GeV. The simulation process is presented briefly in the following, with more details of signal cut-flow being given in Appendix~\ref{sec::appendix}.

\begin{figure*}[thb]
	\centering
		\includegraphics[width=0.4\linewidth]{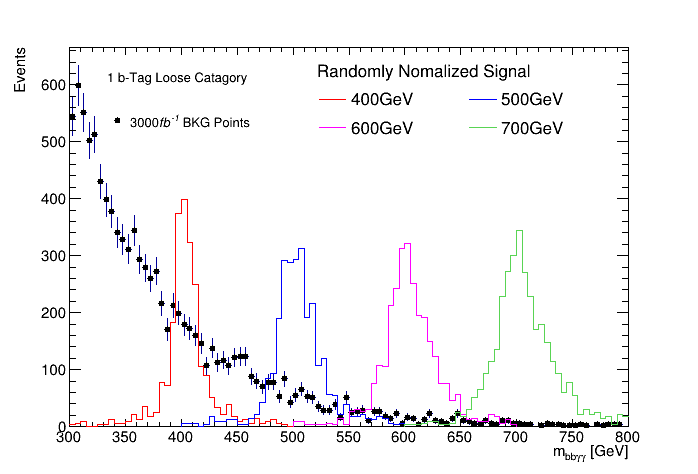}
		\includegraphics[width=0.4\linewidth]{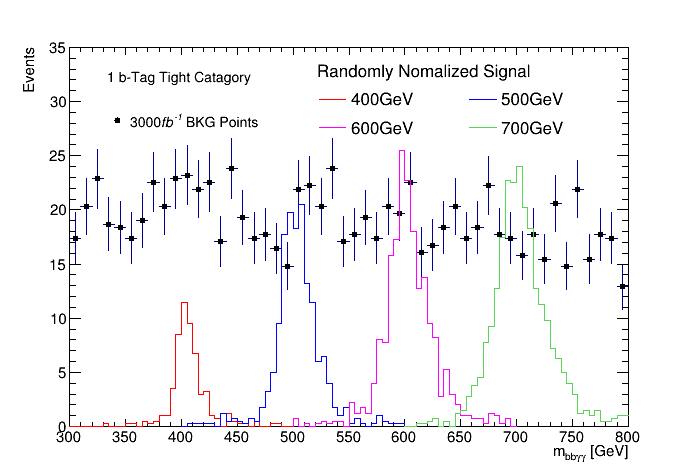}
		\includegraphics[width=0.4\linewidth]{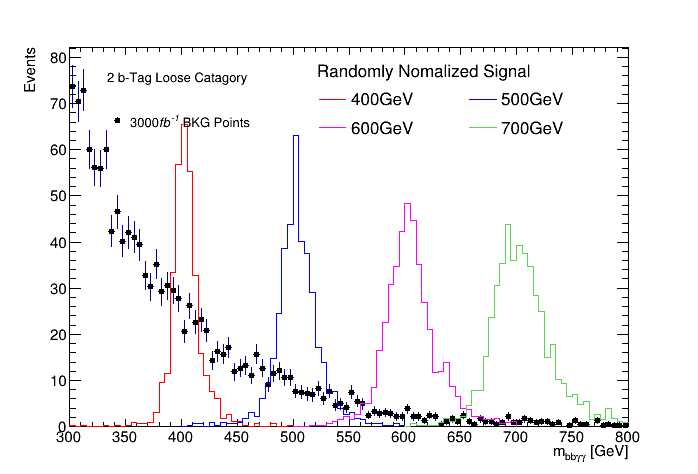}
		\includegraphics[width=0.4\linewidth]{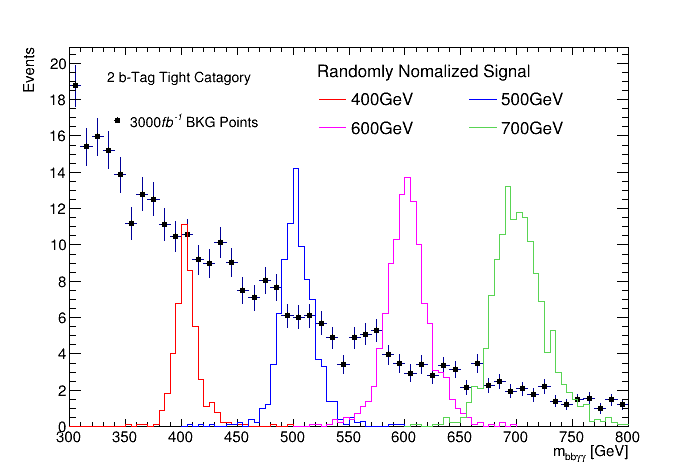}
	\caption{\label{fig::bkg_bbyy}The $m_{bb\gamma\gamma}$ distributions of the $bb\gamma\gamma$ channel analysis, in four signal-enriched regions: ``1-btag loose'', ``2-btag loose'', ``11-btag tight'' and ``2-btag tight'', respectively. The normalization of signal processes are arbitrary.}
\end{figure*}

In the $bb\gamma\gamma$ channel as described in reference~\cite{ATLAS:2018dpp}, the four-body invariant mass $m_{bb\gamma\gamma}$ is used as the discriminating variable. 
Events are categorized into four signal-enriched regions according to number of b-tagged jets as well as loose and tight event selection criteria: "1-btag loose", "2-btag loose", "1-btag tight" and "2-btag tight". The "1-btag" and "2-btag" means that at least one and two jets should be recognized as b-jets.
Acceptances after passing signal-enriched region are in good agreement with the reference analysis, as shown in Table~\ref{tab::cutflow} in Appendix~\ref{sec::appendix}. For background, the $m_{bb\gamma\gamma}$ distributions from reference analysis are extracted. An exponential function is used to fit the $m_{bb\gamma\gamma}$ curve in each signal-enriched region, and then use the fitted function to generate $m_{bb\gamma\gamma}$ distributions according to 3 $ab^{-1}$ luminosity. An $\pm 2\%$ error value is assigned as luminosity calculation uncertainty. The $bb\gamma\gamma$ distributions of signal and background are shown in Fig.~\ref{fig::bkg_bbyy}. The normalizations of signals on the plots are arbitrary. 



\begin{figure}[thb]
	\begin{center}
		\includegraphics[width=0.50\textwidth]{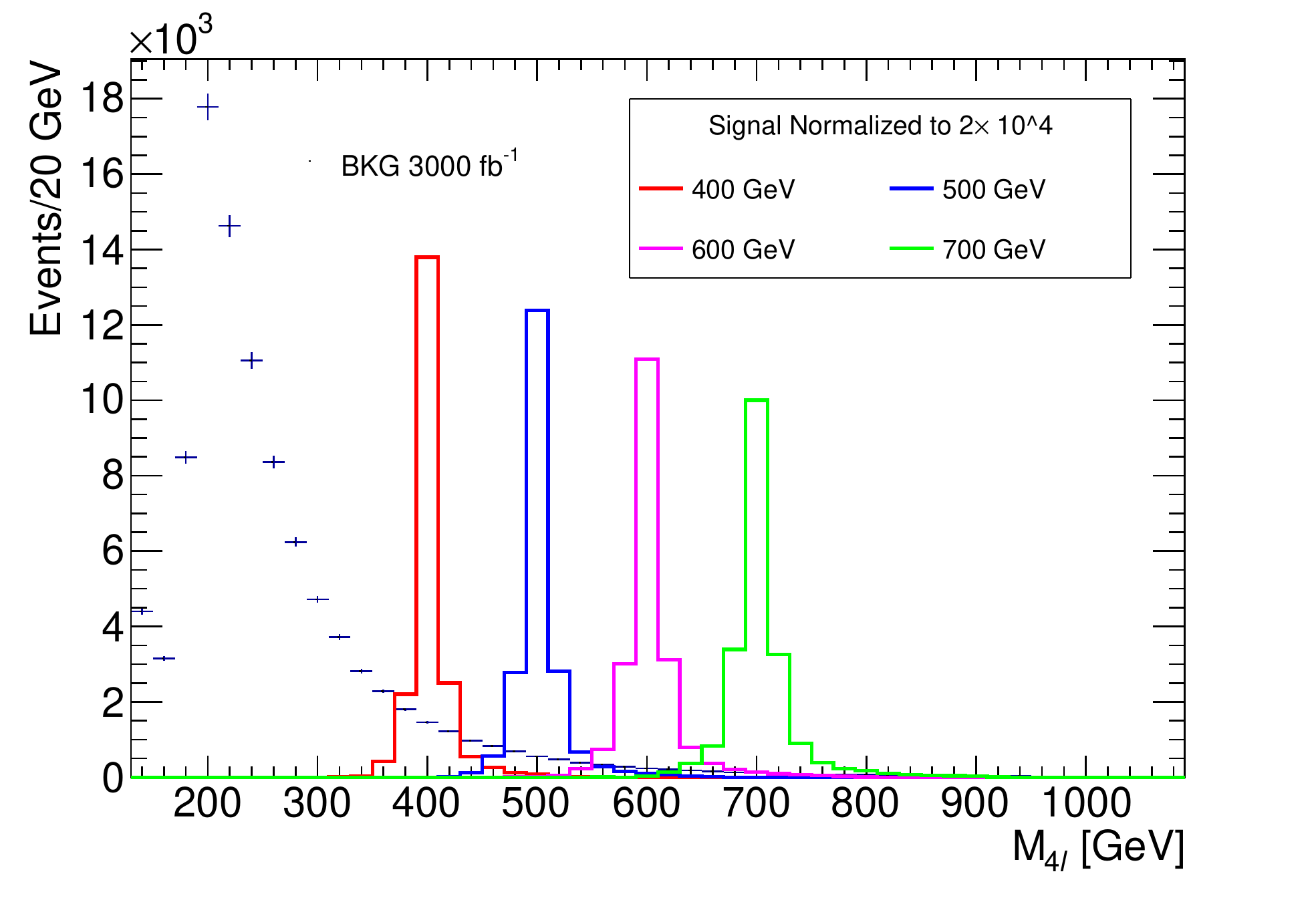}
	\end{center}
	\caption{Distribution of reconstructed four charged leptons for $h_2\to ZZ\to 4\ell$. The number of signal events is normalized to $2\times 10^{4}$. The number of background events is normalized with the integrated luminosity of $3\  ab^{-1}$.}\label{fig:m4l}
\end{figure}

In the $VV\rightarrow 4\ell$ channel, as described in the reference~\cite{ATLAS:2017tlw}, the four-body invariant mass $m_{4\ell}$ is used as the discriminating variable. Events have to pass signal-enriched region selection, which is given with more details in Appendix~\ref{sec::appendix} section B). Signal efficiency is in reasonable agreement with the reference analysis. A similar approach as $bb\gamma\gamma$ analysis is used to simulate background distributions. Fig.~\ref{fig:m4l} shows the $m_{4\ell}$ distributions of signal and background processes. The normalization of signals on the plots are arbitrary.

\begin{figure}[thb]
    \begin{center}
	\includegraphics[width=0.48\textwidth]{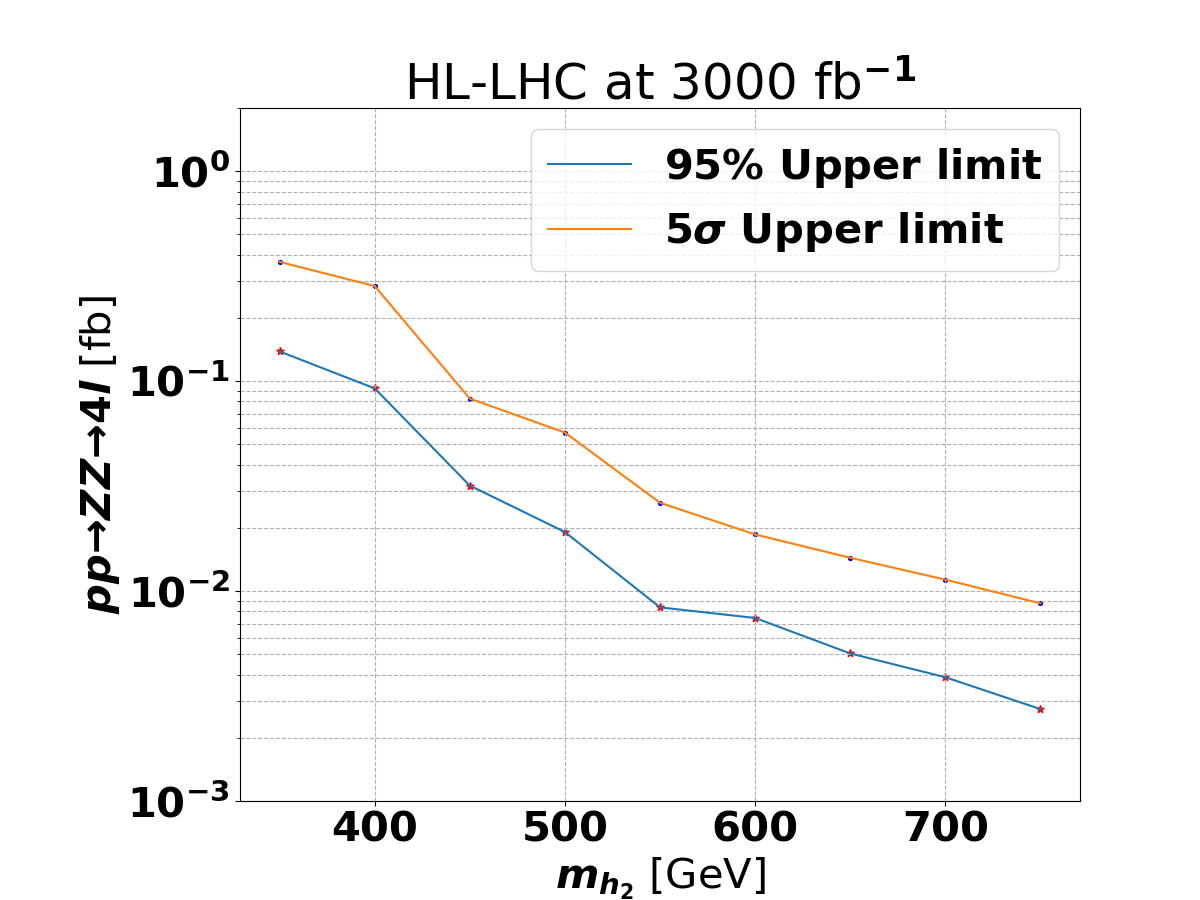}~
        \includegraphics[width=0.48\textwidth]{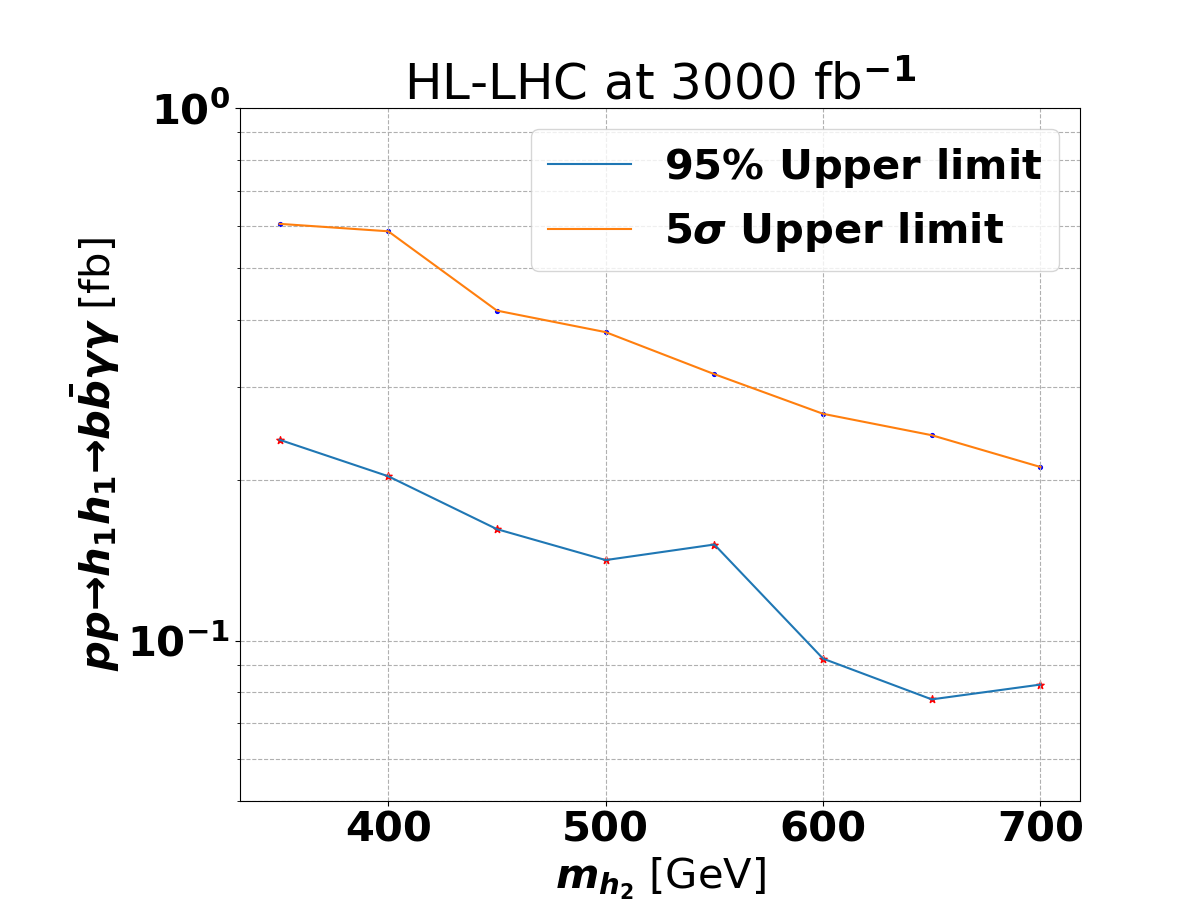}~ 
    \end{center}
    \caption{The HL-LHC prospective 95\% CL$_s$ upper limits and $5\sigma$ discovery limit on the production cross-sections times branching ratios for the two channels. We assume systematic uncertainty for all the backgrounds. }\label{fig:expected-HL}
\end{figure}

With these distributions and efficiencies in hand, we show in Fig.~\ref{fig:expected-HL} the derived the expected 95\% CL$_s$ upper limit for the two individual channels assuming the zero background systematic uncertainties obtained with the python package \texttt{pyhf}~\cite{pyhf,pyhf_joss}. For the discovery limits, we demand that the Gaussian significance converted by the $p_0$ value for the background-only test to be larger than 5  given an observation of the signal plus background predicted in the model in future experiments. In the right plot in Fig.~\ref{fig:expected-HL}, we also plot the discovery limit contours for a combined search for different masses of $h_2$ on the $\sigma_{h_2\rightarrow VV\rightarrow4\ell}$ v.s. $\sigma_{h_2\rightarrow h_1 h_1\rightarrow bb\gamma\gamma}$ plane. These contours are obtained \texttt{pyhf} assuming an uncorrelated signal and background for the two channels. 

\section{Sensitivity and exclusion limits analysis}\label{sec::analysis}

As has been well-studied in earlier work and confirmed in the present analysis, there exists a wide range of xSM parameters compatible with a SFOEWPT. As shown in Fig.~\ref{fig::scan},  the SFOEWPT-compatible choices can lead to non-trivial correlations between $\sigma_{h_2\rightarrow VV\rightarrow4\ell}$ and $\sigma_{h_2\rightarrow h_1 h_1\rightarrow bb\gamma\gamma}$, particularly for the region for smaller cross sections. On the other hand, for the larger values of $\sigma_{h_2\rightarrow VV\rightarrow4\ell}$ this correlation evaporates. 
Generally, for a fixed resonance mass , the  $4\ell$ and $bb\gamma\gamma$ cross sections are determined by $\sin\theta$ and $g_{211}$ respectively. Therefore, for a fixed resonance mass,  $\sigma_{4\ell}$ will vary over a range due to the allowed range for $\sin\theta$. The cross sections would distribute from a minimum and a maximum that are labeled by $\sigma_{4\ell}^{min}$ and $\sigma_{4\ell}^{max}$. Similarly, for a fixed resonance mass  and a fixed $4\ell$ cross section, the $bb\gamma\gamma$ cross section is not determined and locates between the maximam $\sigma_{4\ell;bb\gamma\gamma}^{max}$ and the minimum $\sigma_{4\ell;bb\gamma\gamma}^{min}$.

In order to carry out a systematic analysis of the LHC reach in this parameter space, we adopt the following procedure. 
For each resonance mass point, a set of $\sigma_{h_2\rightarrow VV\rightarrow 4\ell}$ points are selected, which cover the full range between $\sigma_{4\ell}^{min}$ to $\sigma_{4\ell}^{max}$ and distribute uniformly. Then for each fixed $\sigma_{h_2\rightarrow VV\rightarrow 4\ell}$ point, we choose two parameter sets, corresponding to the maximum value , $\sigma_{4\ell;bb\gamma\gamma}^{max}$, and minimum value, $\sigma_{4\ell;bb\gamma\gamma}^{min}$, respectively. 
In total, for each mass point, overall 30$\sim$40 $\sigma_{h_2\rightarrow VV\rightarrow 4\ell}$ points are selected, which results from 60$\sim$80 number of points on the $\sigma_{h_2\rightarrow VV\rightarrow 4\ell}$ and $\sigma_{h_2\rightarrow h_1 h_1\rightarrow bb\gamma\gamma}$ plane.

\begin{figure}[thb]
	\begin{center}
		\includegraphics[width=0.45\textwidth]{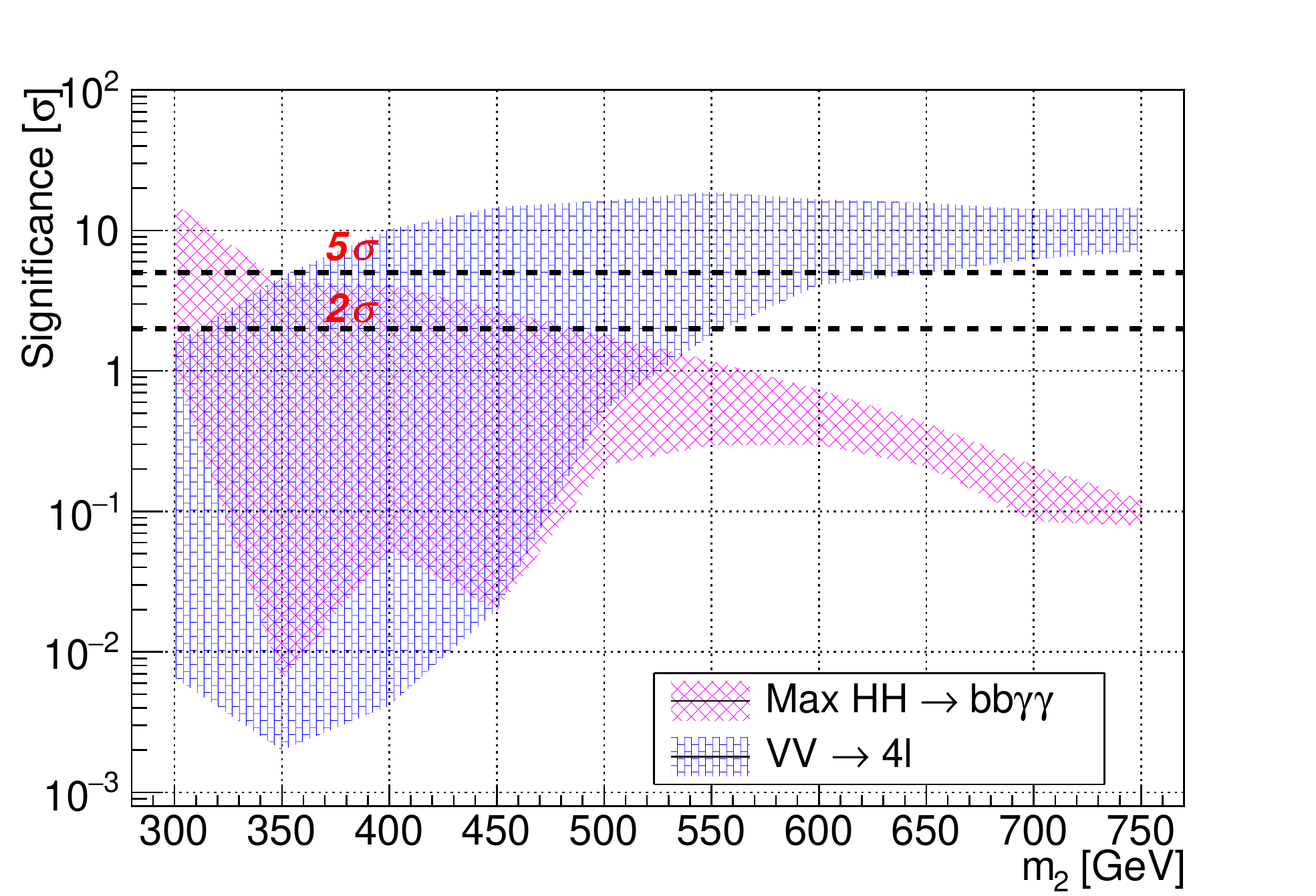}
		\includegraphics[width=0.45\textwidth]{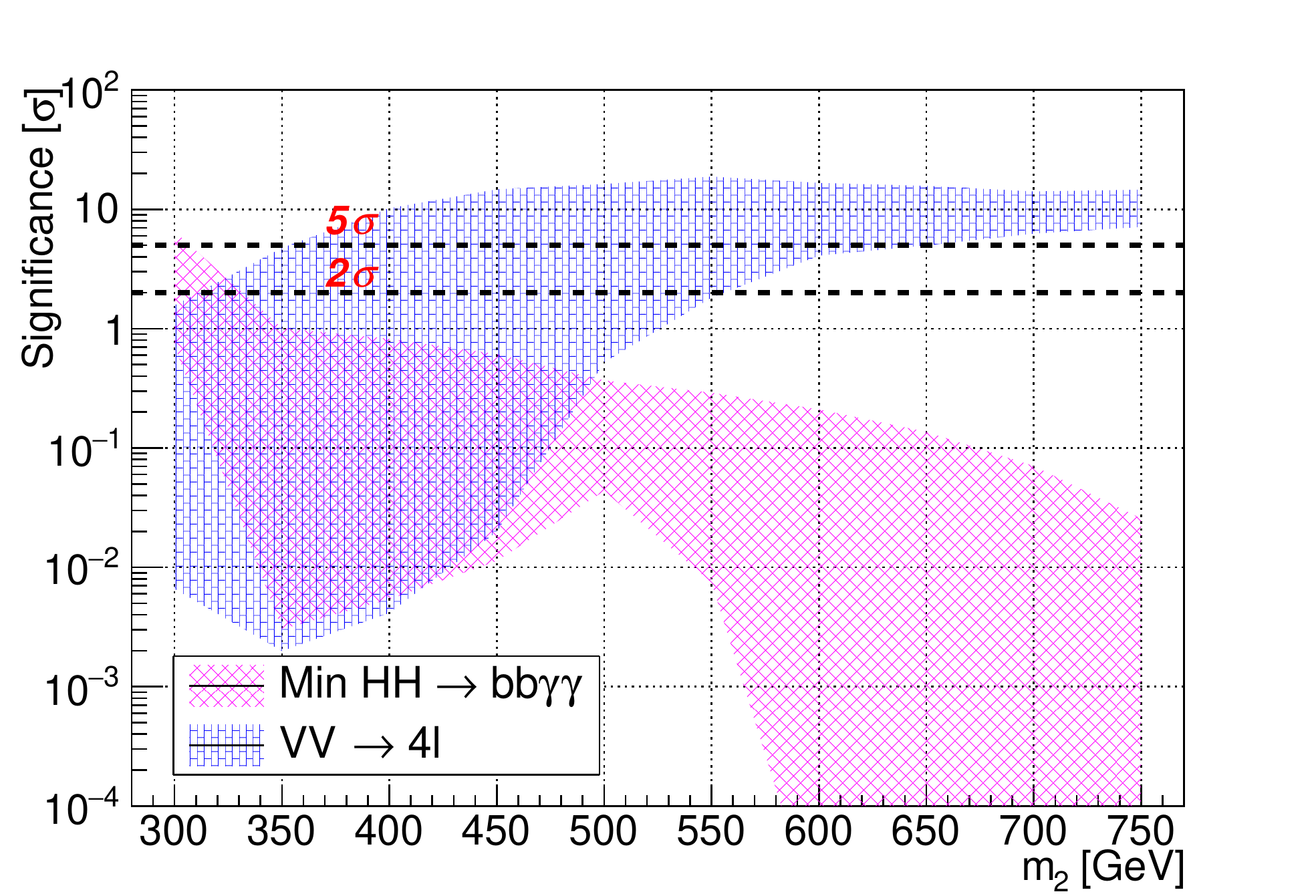}
	\end{center}
	\caption{Prospective discovery/exclusion reach for HL-LHC with 3 $ab^{-1}$. Horizontal and vertical axes give heavy scalar mass and significance, respectively. Red points in the left (right) panels indicate range of maximum (minimum) $\sigma_{h_1 h_1\rightarrow bb\gamma\gamma}$ as one varies over the range of $\sigma_{VV \to 4l}$. The significance for the latter is indicated by the blue points.
	For a fixed $M_S$,  $\sigma_{VV \to 4l}$ depends mainly on $\sin \theta$ mainly. 
	All  points  satisfy the requirement of SFOEWPT.} \label{fig:sigindividulchannel}
\end{figure}

Fig.~\ref{fig:sigindividulchannel} shows the prospective discovery significance for HL-LHC with 3 $ab^{-1}$ integrated luminosity, from the $VV\rightarrow 4\ell$ channel, and from the $h_1 h_1\rightarrow bb\gamma\gamma$ channel with maximum and minimum cross-section, respectively. The $h_1 h_1\rightarrow bb\gamma\gamma$ channel with a maximum cross-section (red region in the left plot) has, on average larger sensitivity than the minimum cross-section (red region in the right plot). Compared to the $h_1 h_1\rightarrow bb\gamma\gamma$, the $VV\rightarrow 4\ell$ channel (blue region) has better sensitivity,  particularly for higher mass resonance signals. The $VV\rightarrow 4\ell$ channel can reach to 5-sigma discovery significance over a wide mass range from 300 GeV to 750 GeV. 

It is worth emphasizing that the $VV\rightarrow 4\ell$ sensitivity is generic for any heavy Higgs that has a $VV$ decay mode and is not specific to the xSM model. Thus, the observation of this mode at a given mass with the significance indicated would be compatible with the SFOEWPT in the xSM but not conclusive. Conversely, the non-observation of this mode would preclude the SFOEWPT-viable xSM with a heavy resonance in a significant portion of parameter space.

\begin{figure}[thb]
	\begin{center}
		\includegraphics[width=0.45\textwidth]{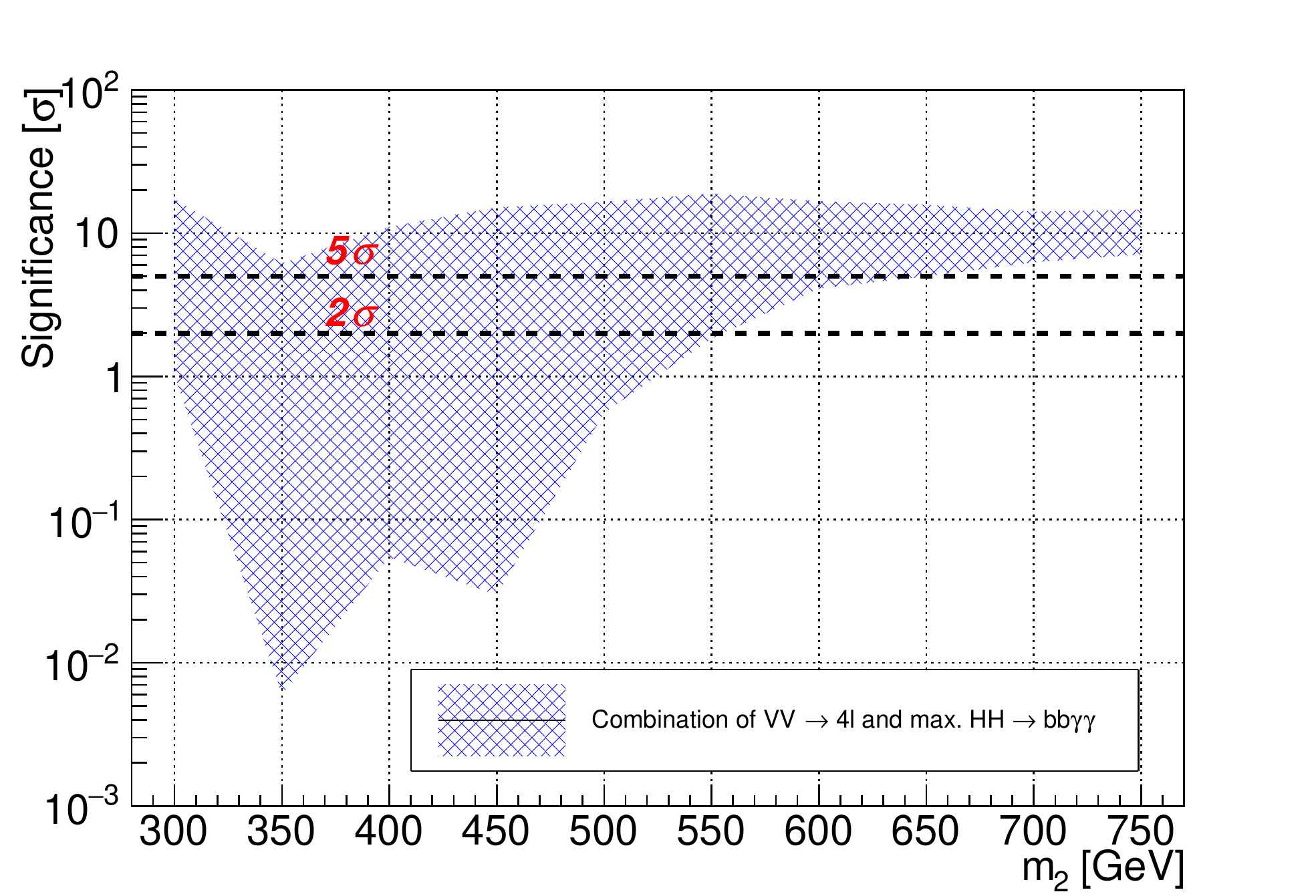}
		\includegraphics[width=0.45\textwidth]{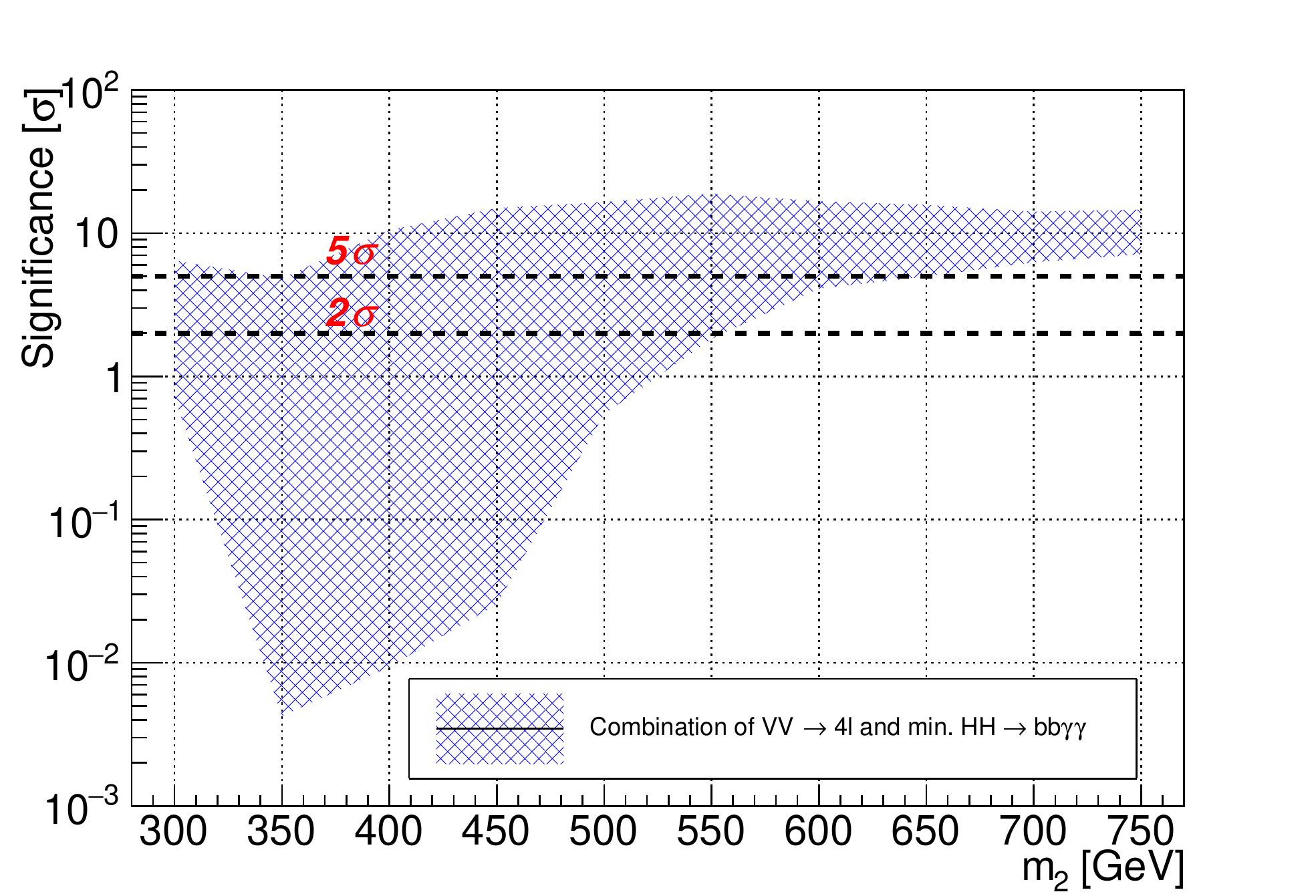}
	\end{center}
	\caption{Prospective discovery/exclusion reach for HL-LHC with 3 $ab^{-1}$
	from the combination of two channels: (1) the $VV\rightarrow 4\ell$ channel, and (2) $h_1 h_1\rightarrow bb\gamma\gamma$ channel with maximum cross section 
	(left panel) and with minimum cross section (right panel).}\label{fig:sigcombination}
\end{figure}

On the other hand, the $h_1 h_1\rightarrow bb\gamma\gamma$ channel would provide a diagnostic probe with some sensitivity up to around 500 GeV. In terms of exclusions corresponding to the 2-sigma line on the plots, both $h_1 h_1\rightarrow bb\gamma\gamma$ and $VV\rightarrow 4\ell$ channels have some sensitivity. In order to enhance the analysis sensitivity, results from the two channels are therefore combined. The combined discovery significance is shown in Fig.~\ref{fig:sigcombination}. More than half of the considered phase space points can be excluded at HL-LHC with 3 $ab^{-1}$ integrated luminosity. On the other hand, there are also points below the 1$\sigma$ line on the plots, which can be further investigated at future colliders.

\begin{figure}[thb]
\centering
    \includegraphics[width=0.48\textwidth]{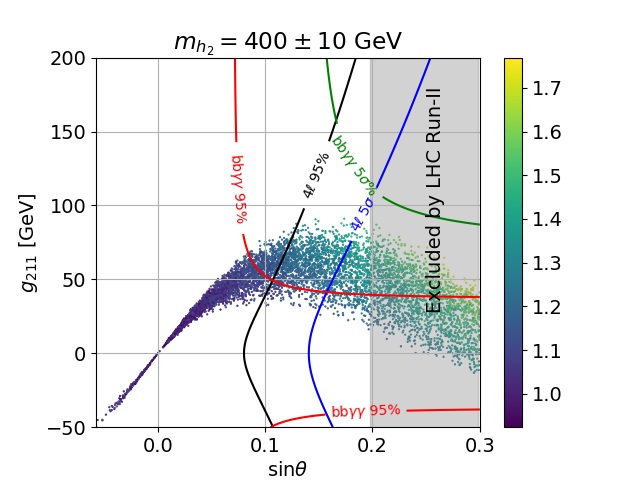}
    \includegraphics[width=0.48\textwidth]{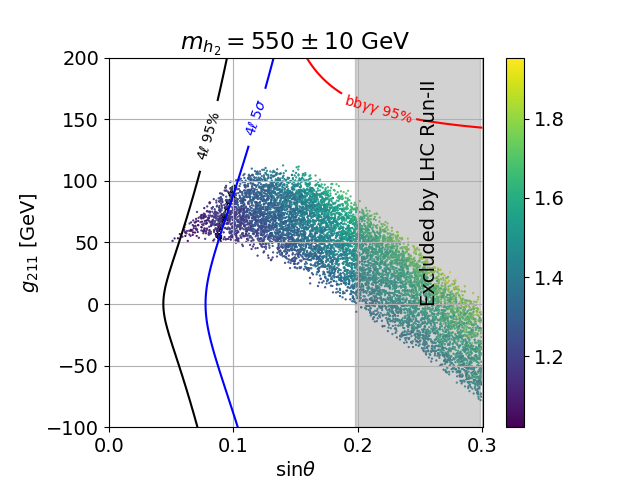}
    \includegraphics[width=0.48\textwidth]{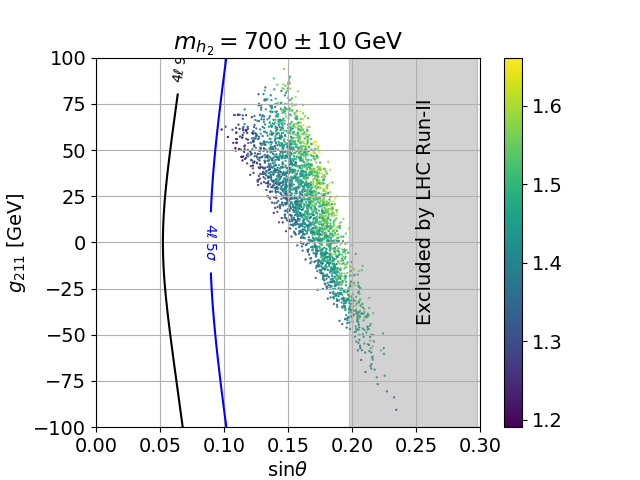}
	\caption{\label{fig::para_add} Prospective HL-LHC (3 $ab^{-1}$) discovery and exclusion reaches for the coupling  $g_{211}/$GeV as a function of $\sin\theta$ allowed by the SFOEWPT  for 400 GeV (the first panel), 550 GeV (the second panel) and 700 GeV (the third panel) resonance masses. Color bars in the panels correspond to the ratio of $g_{111}/g^{\rm SM}_{111}$. The area to the right of each line corresponds to the discovery/exclusion reach for a given channel. The solid lines represent HL-LHC exclusion bound as labeled above 
    The grey region is excluded by LHC Run-II.
    All the points in the plot satisfy theoretical bounds and survive from the EWPT discussed in section.~\ref{subsec::constraints}. }
\end{figure}

One may also interpret the results discovery or exclusion regions in the plane of $g_{211}$ and $\sin\theta$, as shown in Fig.~\ref{fig::para_add}.
From left to right, these plots illustrate the HL-LHC discovery/exclusion potential for --- 400$\pm10$ GeV, 550$\pm10$ GeV, 700$\pm10$ GeV respectively.
All points generate SFOEWPT and satisfy requirements discussed in Sec.~\ref{subsec::constraints}.
Constraints from Higgs measurement at LHC Run \Romannum{2} are labeled by grey shadow. 
Upper limits with 95\% C.L. and $5\sigma$ significance for the $4\ell$ channel are given in all panels.
Only the 95\% C.L. upper limit for the $bb\gamma\gamma$ channel is presented in the first two panels due to its relatively weaker probing ability.
Points located on the right side of each line would be observed/excluded. 
The plots manifest that the $4\ell$ channel shows a much stronger observation/exclusion reach than the $bb\gamma\gamma$ channel, especially for heavy resonance masses.

In more detail: the blue curves and black curves represent the 5$\sigma$ discovery bound and 95\% CL$_s$  exclusion bound for the $ZZ\to 4\ell$ channel. 
The parameter space at the right-hand side of the blue curves with bigger $|\sin\theta|$ and $|g_{211}|$ can be discovered with 5$\sigma$ significance at HL-LHC if the nature is realized by the real singlet model, and the parameter spaces at the right-hand side of the black curves can be excluded with non-observation of the signal.  The green and red curves are for 5$\sigma$ discovery bound and 95\% CL$_s$  exclusion bound for the $h_1h_1\to bb\gamma\gamma$ channel.  All these bounds assume an integrated luminosity of 3000 fb$^{-1}$. In the plot for $m_{h_2} = 700$ GeV, only the blue, black and red curves exist in the parameter space of interest because the $bb\gamma\gamma$ has no detection capability for this mass region. 
The shaded regions in each plot are excluded by current LHC Run-II results.

\begin{figure}[thb]
	\begin{center}
  		\includegraphics[width=0.32\textwidth]{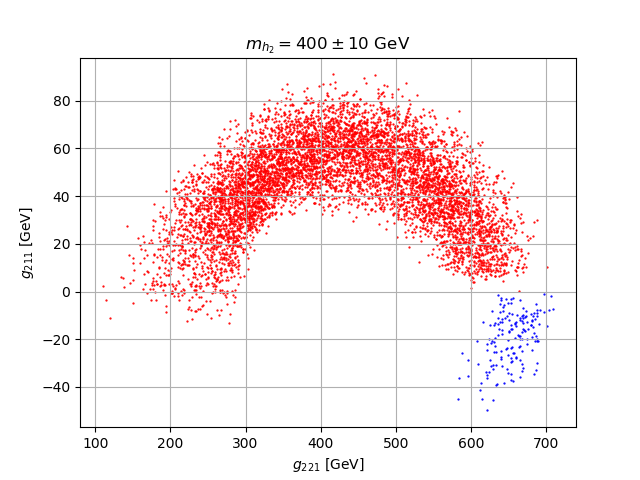}
		\includegraphics[width=0.32\textwidth]{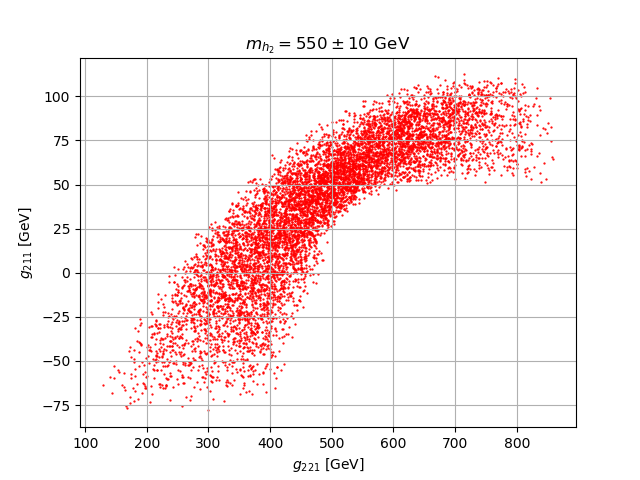}
		\includegraphics[width=0.32\textwidth]{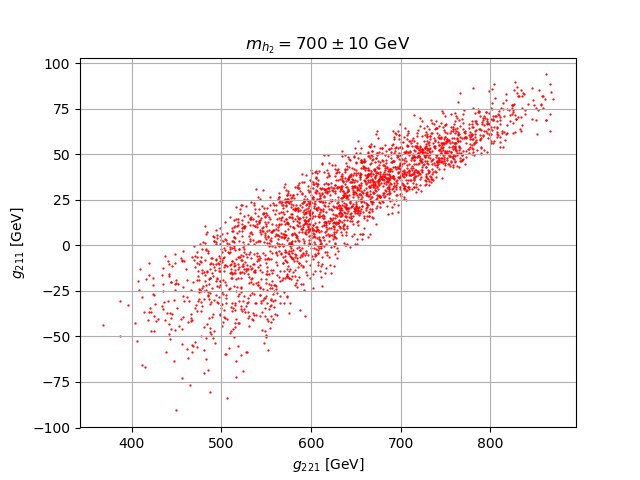}
	\end{center}
	\caption{\label{fig::g211_g221}  Points in the $g_{211}$ vs $g_{221}$ plane, where we have removed the points that are excluded by current LHC measurements. The red (blue) color indicates that the sign of $\sin\theta$ for the corresponding point is positive (negative). }
\end{figure}

With these results in mind, one can anticipate several possible experimental outcomes:

\begin{itemize}
\item[(i)]Both modes are observed:
If both modes with a resonance mass  $\lesssim 500$ GeV (Fig.~\ref{fig:sigindividulchannel}) are observed by 2$\sigma$, or the 4$\ell$ channel reaches 5$\sigma$ with the $bb\gamma\gamma$ reaching 2$\sigma$, it is expected by, and shows strong evidence for, the SFOEWPT-viable xSM.
However, observation of both channels by 5$\sigma$ would be inconsistent with the SFOEWPT-viable xSM except for $m_{h_2}\sim 350$ GeV. 
            
\item[(ii)] Neither channel is observed at the HL-LHC: The non-observation of both the $4\ell$ and di-Higgs channels does not rule out the SFOEWPT-viable xSM. However, it implies an intriguing pattern regarding the sign correlation between $\sin\theta$ and $g_{211}$. Notably, for parameter points remaining unexcluded by the future HL-LHC experiment, a negative $\sin\theta$ is only consistent with a relatively light $h_2$ mass. Additionally, in cases of negative $\sin\theta$, there is a strong correlation between $g_{211}$ and $\sin\theta,$ as indicated by the narrow strip in the left plot of Fig.~\ref{fig::g211_g221}. We discuss this correlation and Fig.~\ref{fig::g211_g221} in more detail below.
\item[(iii)] The $VV\rightarrow 4\ell$ is observed but the  $h_1 h_1\rightarrow bb\gamma\gamma$ is not: If the observed resonance satisfies $\gtrsim$ 350 GeV, the existence of a heavy resonance compatible with the SFOEWPT-xSM would be established but a future more sensitive search for the $h_1 h_1\rightarrow bb\gamma\gamma$ would be needed. 
On the other hand, if the resonance is $\lesssim$ 350 GeV, the observation would be in consistent with xSM prediction and is hence ruled out.
\item[(iv)] The $h_1 h_1\rightarrow bb\gamma\gamma$ is observed but the $VV\rightarrow 4\ell$ is not: 
If one observes the di-Higgs channel with $bb\gamma\gamma$ final states in the small resonance mass region ($\lesssim$ 350 GeV) and no significant observations for the $4\ell$ channel, the result would be consistent with the SFOEWPT in the xSM. 
On the contrary, if the observation occurs with heavier-mass resonance ($\gtrsim$ 500 GeV), the SFOEWPT-viable xSM prediction would be inconsistent with the observation and, therefore, the xSM FOEWPT would be ruled out.
\end{itemize}
Note that for each possible outcome, one would need to perform an updated xSM parameter scan that takes into account the various experimental results.


We now comment further on the question of the sign of  $\sin\theta$ and the correlation with other xSM tri-scalar couplings. 
First, we observe that the independent measurement of the sign of $\sin\theta$ is not an easy task, and one cannot determine the sign of $\sin\theta$ with solely the couplings between $h_2$ and SM fermions or vector bosons, because in this case each propagator of $h_2$ present in a diagram must contribute to 2 powers of $\sin\theta$ erasing the sign information.  
In this case, the sign of the $\sin\theta$ must be resolved by a global fit that includes physical processes that are sensitive to all the scalar couplings, such as $g_{221}$. As an illustration, in Fig.~\ref{fig::g211_g221}, we show the scatter plot in the $g_{211}$ vs 
$g_{221}$ plane for all the points that can give SFOEWPT and simultaneously satisfy all the theoretical and current experimental bounds, including the one from LHC run-II. The color of the points indicates the sign of the mixing angle $\sin\theta$, red for positive and blue for negative. 
One can find that for $m_{h_2}$ around 400 GeV, the points with the different sign of $\sin\theta$ tend to populate different parameter space in this 2-D plane, which indicates that the measurement of the two tri-linear couplings could help to disentangle the sign of $\sin\theta$ in xSM.

\section{Conclusion}\label{sec::conclusion}
	Exploring the thermal history of electroweak symmetry breaking is an important effort in its own right and may yield clues to solving the baryogenesis problem through EWBG.
	This scenario requires a SFOEWPT as a pre-condition for generation of the CPV asymmetries that ultimately yield the baryon asymmetry.
	In the SM,  EWSB occurs though a cross-over transition.
	However,  the Higgs portal interactions in the xSM  can generate a SFOEWPT for suitable choices of model parameters. 
    The portal interactions are strongly correlated with the strength of the EWPT. 
    Generally, a successful SFOEWPT requires  relatively large portal interactions.
    However, constraints from Higgs measurement, EWPO and perturbativity indicate that the magnitudes of the mixing angle and the portal couplings cannot be arbitrarily large. 
    Therefore, the parameter space is highly constrained. 
    Through our analysis, we find that the remained parameter space has different sensitivies for different detection channels.
    As is well-known, one of the effective ways to detect the Higgs portal interactions is the heavy scalar resonance search via di-Higgs channel.
    For the di-Higgs channel, we have considered in particular the $bb\gamma\gamma$ final state.
    Another way to probe the Higgs portal interactions is through the di-boson channel. For the di-boson channel, we have considered in particular the $4\ell$ final state. A novel aspect of this study is the analysis of the complementarity between these two channels and the discovery/exclusion reach that results from their combination.

    Through our analysis We find that the $b\bar{b}\gamma\gamma$ channel is able to explore the low mass region with $m_2 \lesssim 320$ GeV effectively and the $4\ell$ channel is more suitable for the high mass region with $m_2 \gtrsim 500$ GeV.
    It is also possible that a combination of the two search channels can push the significance above 5$\sigma$ even though the significance for each individual channel is less than 5$\sigma$.
In terms of exclusion, one is able to exclude the portion of SFOEWPT parameter space by taking advantage of the combination of $b\bar{b}\gamma\gamma$ and $4\ell$ channels over roughly $m_2 \lesssim 750$ GeV at the 14 TeV HL-LHC with luminosity 3$ab^{-1}$.
Moreover, the combination of both channels provides an important diagnostic probe of the SFOEWPT-viable xSM. However, signal events associated with minimal $b\bar{b}\gamma\gamma$ cross section are difficult to exclude at the HL-LHC, pointing to the need for  a future collider with higher energy to obtain a comprehensive probe (see, {\it e.g.}, Refs.~\cite{Kotwal:2016tex,Ramsey-Musolf:2019lsf}). 
Finally, in the case of combined discovery, there is a strong correlation between $g_{211}$ and $\sin\theta$. On the other hand, even though a non-observation of both channels would not rule out SFOEWPT-viable xSM, it would put constraints on the sign of the mixing angle and the tri-linear scalar coupling $g_{211}$. 
In this case, future high-energy collider experiments may needed to further diagnose and probe the different scalar couplings in the model (see, {\it e.g.}, Refs.~\cite{Kotwal:2016tex,Ramsey-Musolf:2019lsf}). We show as an example how the simultaneous measurement of $g_{211}$
and $g_{221}$ could allow one to disentangle the sign of $\sin\theta$ in the SFOEWPT-viable xSM.

\section{Acknowledgement}
SA, MJRM and WZ are supported in part under National Natural Science Foundation of China grant No. 11975150. H.-L.L. is supported by F.R.S.-FNRS through the IISN convention "Theory of Fundamental Interactions” (N: 4.4517.08). KL is supported by the Major International (Regional ) Joint Research Program of NSFC-CERN (Grant No.12061141004).

\section{Appendix}\label{sec::appendix}

\subsection{Two-loop $\beta$-functions for the dimensionless parameters}\label{sec::appdxrge}
We use the python package \texttt{PyR@TE 3}~\cite{Sartore:2020gou} to derive the renormalization group equations (RGEs) in the $\overline{{\rm MS}}$ scheme at one-loop level and checked that the results in the limiting cases are in agreement with those in literatures ~\cite{Schienbein:2018fsw, Luo:2002ti, Luo:2002ey, Gonderinger:2009jp}.
The full one-loop RGEs in the form of $ {\rm d}X/{\rm d}\log\mu = \beta^{(1)}(X)/(4\pi)^2$ are given as follows: 
\begin{eqnarray}
    \beta^{(1)}(g_1) &=&\frac{41}{6} g_1^{3}\\
    \beta^{(1)}(g_2) &=&- \frac{19}{6} g_2^{3}\\
    \beta^{(1)}(g_3) &=&-7 g_3^{3}\\
    \beta^{(1)}(y_t) &=&
+ \frac{9}{2} y_t \left|{y_t}\right|^{2}
+ \frac{3}{2} y_t \left|{y_b}\right|^{2}
-  \frac{17}{12} g_1^{2} y_t
-  \frac{9}{4} g_2^{2} y_t
- 8 g_3^{2} y_t\\
\beta^{(1)}(y_b) &=&
+ \frac{9}{2} y_b \left|{y_b}\right|^{2}
+ \frac{3}{2} y_b \left|{y_t}\right|^{2}
-  \frac{5}{12} g_1^{2} y_b
-  \frac{9}{4} g_2^{2} y_b
- 8 g_3^{2} y_b\\
\beta^{(1)}(\lambda) &=&
+ 24 \lambda^{2}
+ \frac{1}{2} a_{2}^{2}
- 3 g_1^{2} \lambda
- 9 g_2^{2} \lambda
+ \frac{3}{8} g_1^{4}
+ \frac{3}{4} g_1^{2} g_2^{2}
+ \frac{9}{8} g_2^{4}
\nonumber \\
&&+ 12 \lambda \left|{y_t}\right|^{2}
+ 12 \lambda \left|{y_b}\right|^{2}
- 6 \left|{y_t}\right|^{4}
- 6 \left|{y_b}\right|^{4}\\
\beta^{(1)}(a_{2}) &=&
+ 12 a_{2} \lambda
+ 4 a_{2}^{2}
+ 6 a_{2} b_{4}
-  \frac{3}{2} a_{2} g_1^{2}
-  \frac{9}{2} a_{2} g_2^{2}
+ 6 a_{2} \left|{y_t}\right|^{2}
+ 6 a_{2} \left|{y_b}\right|^{2}\\
\beta^{(1)}(b_{4}) &=&
+ 2 a_{2}^{2}
+ 18 b_{4}^{2}\\
\beta^{(1)}(a_{1}) &=&
-  \frac{3}{2} a_{1} g_1^{2}
-  \frac{9}{2} a_{1} g_2^{2}
+ 12 a_{1} \lambda
+ 4 a_{1} a_{2}
+ 4 a_{2} b_{3}
\nonumber \\
&&+ 6 a_{1} \left|{y_t}\right|^{2}
+ 6 a_{1} \left|{y_b}\right|^{2}
+ 2 a_{1} \left|{y_\tau}\right|^{2}\\
\beta^{(1)}(b_{3}) &=&
+ 3 a_{1} a_{2}
+ 18 b_{3} b_{4}\\
\beta^{(1)}(\mu^2) &=&
-  \frac{3}{2} g_1^{2} \mu^2
-  \frac{9}{2} g_2^{2} \mu^2
-  \frac{1}{2} a_{1}^{2}
+ 12 \lambda \mu^2
-  a_{2} b_{2}\nonumber\\
&&+ 6 \mu^2 \left|{y_t}\right|^{2}
+ 6 \mu^2 \left|{y_b}\right|^{2}
+ 2 \mu^2 \left|{y_\tau}\right|^{2}\\
\beta^{(1)}(b_{2}) &=&
+ a_{1}^{2}
+ 4 b_{3}^{2}
- 4 a_{2} \mu
+ 6 b_{2} b_{4},
\end{eqnarray}
where $g_1,g_2,g_3$ are $U(1)_y$, $SU(2)_L$ and $SU(3)_c$ gauge couplings respectively and $y_t$ and $y_b$ are Yukawa couplings for top and bottom quarks. Their definitions are summarized in the following equations:
\begin{eqnarray}
   && D_\mu \Psi = (\partial_\mu-ig_1q B_\mu -i g_2T_2^{i}W^i_\mu-ig_3T_3^a G^a) \Psi,\\
    &&{\cal L}_{\rm Yukawa} = y_t\bar{Q}_L \tilde{H} u_R + y_b \bar{Q}_L H d_R + h.c.,
\end{eqnarray}
where $q$ is the $U(1)_y$ charge of the field $\Psi$, $T_2$ and $T_3$ are the corresponding generators of the $SU(2)$ and $SU(3)$ groups depending on the representations of the field $\Psi$. For example, for the quark left-handed doublet $Q_L$, it has $q=1/6$, $T^2_2=\sigma^i/2$ and $T_3^a=\lambda^a/2$ with $\sigma^i$ and $\lambda^a$ the Pauli and Gell-Mann matrices respectively. $\tilde{H}_i\equiv i\sigma^2_{ij} H_j$, $u_R$ and $b_R$ are right-handed up-type and down-type quark singlets respectively.

For the initial inputs of SM Yukawa and gauge couplings $g_1,g_2,g_3, y_t,y_b$, we use the package \texttt{mr}~\cite{Kniehl:2016enc} to convert the input parameters from the On-Shell scheme to the $\overline{\rm MS}$ scheme and run to the scale $\mu = 246$ GeV. 
When converting between the On-shell and $\overline{\rm MS}$ scheme, we use the relations in the order of one-loop QCD and EW in the Standard model and assume that the correction from the heavy Higgs is small provided that the $\sin\theta$ is within the constraint of EWPO.
For the running couplings $g_1,g_2,g_3, y_t,y_b$, we implement the one-loop RGEs for consistency, which is not affected by the heavy singlet in xSM. 
The On-Shell input parameters are obtained from the Particle Data Group~\cite{ParticleDataGroup:2022pth}:
\begin{eqnarray}
    &&M_b = 4.78~{\rm GeV}, ~M_t = 172.69~~{\rm GeV}, ~ M_W = 80.377~{\rm GeV}, ~ M_Z = 91.1876~{\rm GeV}, \\
    && M_h = 125.25~{\rm GeV}, ~ G_F = 1.16638~{\rm GeV^{-2}}, ~\alpha_s(M_t) = 0.108102,
\end{eqnarray}
from which we obtain the $\overline{\rm MS}$ couplings at the scale $\mu=246$ GeV as follows:
\begin{eqnarray}
    &&g_1 = 0.36001, ~ g_2 = 0.64632, ~ g_3 = 1.15330, \\
    && y_t = 0.93930, ~ y_b = 0.01897.
\end{eqnarray}

\subsection{bb$\gamma\gamma$ channel simulation}\label{sec::appdx1}
We perform a generation of a signal process $p p \to h_2 \to h_1 h_1 \to b\bar{b} \gamma\gamma$ by utilizing MadGraph5~\cite{Alwall:2011uj} in parton level and use Pythia8~\cite{Sjostrand:2007gs} to simulate the parton shower process followed by a detector simulating software Delphes3~\cite{deFavereau:2013fsa} to simulate the detector response of this process at LHC. In this section, we will follow the ATLAS analysis in Ref.~\cite{ATLAS:2018dpp} and reproduce the cutflow efficiency compared with the experimental data. 

In the detector simulation, photon candidates must satisfy isolation criteria of $\Delta R \leq 0.2$ for both calorimeter- and track-based isolation. In track-based isolation, the transverse momentum within the cone ($p_T^{iso}$) counts only the tracks with $p_T > 1$ GeV. 
Jets are reconstructed using anti-$k_t$ clustering algorithm~\cite{Cacciari:2008gp} with a radius parameter set to be $R=0.4$ and are required to satisfy $|\eta|<2.5$ and $p_T > 25$ GeV. In addition, the efficiency for a b-quark jet to pass the $b-$tagging requirement is set to be a constant equal to 70\%~\cite{CMS:2012feb} approximately. 

The selection is simplified according to Ref.~\cite{ATLAS:2018dpp}:

\begin{itemize}
	\item Events are required to have at least two photons and two jets with one or two jets recognized as b-jet(s). Any event containing more than two b-jets is rejected. 
	\item The invariant mass of two photon with the highest $p_T$ should satisfy 105 GeV $< m_{\gamma\gamma} < 160$ GeV.
	\item For each photon, it is required to have $E_T/m_{\gamma\gamma} > 0.35$ and 0.25 respectively.
	\item Four signal regions are defined based on the highest-$p_T$ and next-highest $p_T$(They are called "Cen Jets" in Table.~\ref{tab::cutflow}). If $p_T^1 > 40$GeV and $p_T^2 > 25$GeV, the signal is labeled with "loose". If $p_T^1 > 100$GeV and $p_T^2 > 30$GeV, the signal is labeled "tight". Combining with the b-jet number, signal events are classified to four regions, which are "1-btag loose", "1-btag tight", "2-btag loose" and "2-btag tight".
	\item For tight (loose) signal regions, the diphoton invariant mass is required to be 125 GeV$\pm$ 4.3(4.7) GeV.
	\item The dijet system is rescaled by a factor of $m_H/m_{jj}$.
	\item Loose selection is used for resonances with $m_X \leq 400$ GeV and the tight selection for resonances with $m_X \geq 400$ GeV. For tight (loose) selection, we require 335 GeV $< m_{\gamma\gamma j j} <$ 1140 GeV(245 GeV $< m_{\gamma\gamma j j} <$ 610 GeV).
	\item For events that pass the selection above, we obtain their $m_{bb\gamma\gamma}$ distribution.
\end{itemize}

\begin{table*}[htbp]\centering
	\begin{tabular}{|c|c|c|c|c|c|c|} \hline
		& $m_{\gamma \gamma}$ & 2 Cen Jets & b-tagging & bjet pT & $m_{bb}$ & $E_T/m_{\gamma\gamma}$   \\ \hline
		Exp(300GeV 1-btag)  & 37.6\% & 27.80\% & 11.8\% & 10.8\% & 5.9\%  & 5.6\%        \\ \hline
		Sim(300GeV 1-btag)  & 37.6\% & 27.9\% & 12.65\% & 9.56\% & 5.52\%  & 5.39\%   \\ \hline
		Exp(300GeV 2-btag)  & 37.6\% & 27.8\% & 8.1\% & 7.9\% & 6.8\%  & 6.4\% \\\hline	
		Sim(300GeV 2-btag)  & 37.6\% & 27.9\% & 7.63\% & 7.09\% & 5.84\%  & 5.82\%  \\\hline	
		Exp(400GeV 1-btag)  & 40.2\% & 32.8\% & 14.3\% & 5.5\% & 2.5\%  & 2.4\%        \\ \hline
		Sim(400GeV 1-btag) & 40.2\% & 32.7\% & 13.44\% & 5.42\% & 1.98\% & 1.98\%  \\ \hline
		Exp(400GeV 2-btag)  & 40.2\% & 32.8\% & 10.3\% & 4.3\%  & 3.6\%  & 3.4\%  \\\hline	
		Sim(400GeV 2-btag)  & 40.2\% & 32.7\% & 12.05\% & 4.85\%  & 3.70\%  & 3.70\% \\\hline
		
	\end{tabular}
	\caption{\label{tab::cutflow} Cutflows for higgs production at LHC with $\sqrt{s}=1.3$ TeV, $m_{h_2}=300/400$ GeV, $\mathcal{L}=250fb^{-1}$. The  Exp-line shows the experimental result in ref~\cite{ATLAS:2018dpp}. The Sim-line shows the simulation result.}
\end{table*}

\subsection{$ZZ\to 4\ell$ channel simulation and selection}\label{sec::appdx2}
We generate the parton level signal events $pp\to h_2\to ZZ\to \ell^+\ell^-\ell^+\ell^-$ with MadGraph5~\cite{Alwall:2011uj}, then showering with Pythia8~\cite{Sjostrand:2007gs} and simulate the detector effect with Delphes3~\cite{deFavereau:2013fsa}. We follow the ATLAS analysis in Ref.~\cite{ATLAS:2017tlw}, and reproduce similar signal and background efficiencies in that paper.
The analysis cut flow is described as followings:
\begin{itemize}
    \item Select the events with two same-flavor opposite-sign lepton pairs. The reconstructed electron (muon) must have $p_T>7(5)$ GeV and $|\eta|<2.47 (2.7)$.
    \item For lepton pairs, we denote the invariant mass of the pair closer to $Z$ mass as $m_{12}$ and that of the other pair as $m_{34}$. We demand that $m_{12}$ and $m_{34}$ satisfying the following cuts:  
    \begin{eqnarray}
    &&50\ {\rm GeV}< m_{12} < 106\ {\rm GeV}\nonumber \\
    &&m_{34}<116\ {\rm GeV}\nonumber \\
    &&m_{34}>\begin{cases*}
                   12\ {\rm GeV} & $m_{4\ell}\leq 140\ {\rm GeV}$  \\
                   12+50\frac{m_{4\ell}-140}{190-140} & $140\ {\rm GeV}<m_{4\ell}<190\ {\rm GeV}$\\
                   50\ {\rm GeV}  & $m_{4\ell}\geq 190\ {\rm GeV}$
                 \end{cases*}
    \end{eqnarray}
    \item We also require that leptons are separated with each other by $\Delta R>0.1$ if they are the same flavor, and $\Delta R>0.2$ otherwise.
    \item For $4\mu$ and $4e$ events, we veto the events containing opposite-sign lepton pairs with $m_{\ell \ell}<5\ {\rm GeV}$.
    \item We implement the cut on the track-isolation discriminant as the sum of the transverse momenta of tracks within $\Delta R=0.3$ (0.2) of the muon (electron) candidate excluding the lepton track, divided by the $p_T$ of the lepton. Such discriminants are required to be smaller than 0.15.  
\end{itemize}
Compared to Ref.~\cite{ATLAS:2017tlw}, we did not impose the calorimeter-based isolation requirement because they are technically difficult to implement in the Delphes. In practice,, we find that the major effect on the signal efficiencies is from the detector efficiencies for leptons, and with the above cuts we are already able to obtain signal efficiencies reasonably close to those in the ATLAS paper~\cite{ATLAS:2017tlw}. In table.~\ref{tab::4leff}, we illustrate the effectiveness of our simulation by showing the comparison of signal efficiencies for different lepton channels for a fixed benchmark point $m_{h_2}=600\ {\rm GeV}$, we only compare the final efficiencies because the efficiencies for the intermediate cuts are inaccessible in the experimental paper.
\begin{table*}[htbp]\centering
	\begin{tabular}{|c|c|c|} \hline
	 &Sim (600 GeV) & Exp (600 GeV)\\
	 \hline
	$4\mu$ & 42.2\% & 48\% \\
	\hline
	$4e$ & 63.4\% & 64\%\\
	\hline
	$2\mu 2e$ & 51.7\%& 57\%\\
	\hline
	\end{tabular}
	\caption{\label{tab::4leff} Signal efficiencies for different lepton channels of the process the $h_2\to ZZ\to 4\ell$, the Exp numbers are obtained from Ref.~\cite{ATLAS:2017tlw}. }
\end{table*}
\phantomsection

\subsection{Comments on the constraint from  electroweak precision observables (EWPO)}\label{sec::new_ewpo}
xSM can influence  EWPO in two ways: rescaling couplings between the SM-like Higgs to SM particles and generating new diagrams involving the single-like Higgs. When parameterizing theoretical predictions of  EWPO in the framework of the oblique parameters  $S,T$ and $U$~\cite{Peskin:1990zt,Peskin:1991sw}, the shifted oblique parameters in the xSM can be uniquely determined by the two physical parameters: the mixing angle $\theta$ and the mass of the singlet-like Higgs $m_{2}$, which can be expressed in the following formula,
\begin{eqnarray}
\Delta {\cal O}=(\cos^2\theta-1){\cal O}^{\rm SM}(m_{h_1})+\sin^2\theta {\cal O}^{\rm SM}(m_{h_2})=\sin^2\theta \left[{\cal O}^{\rm SM}(m_{h_2})-{\cal O}^{\rm SM}(m_{h_1})\right],
\end{eqnarray}
where ${\cal O}$ can be either $S,T$ or $U$, and ${\cal O}^{\rm SM}(m_{h_2})$ are the SM expression for ${\cal O}$ if the Higgs mass were to be $m_{h_2}$. Therefore a constraint on the $\cos\theta-m_{h_2}$ plane can be derived by constructing the $\chi^2$ with the predicted oblique observables and the experimental data. In our previous work~\cite{Huang:2017jws}, we used the global fitted center values, uncertainties, and correlations for these oblique parameters from the Gfitter group~\cite{Baak:2014ora} and obtained a lower bound on  $\cos\theta$ as a function of $m_{h_2}$. However, the recently updated measurement of $m_W$ from the CDF collaboration~\cite{CDF:2022hxs} exhibits a significant deviation from the SM prediction, which indicates a qualitative change in the EWPO constraint for xSM. 

\begin{figure}[thb]
	\begin{center}
		\includegraphics[width=0.5\textwidth]{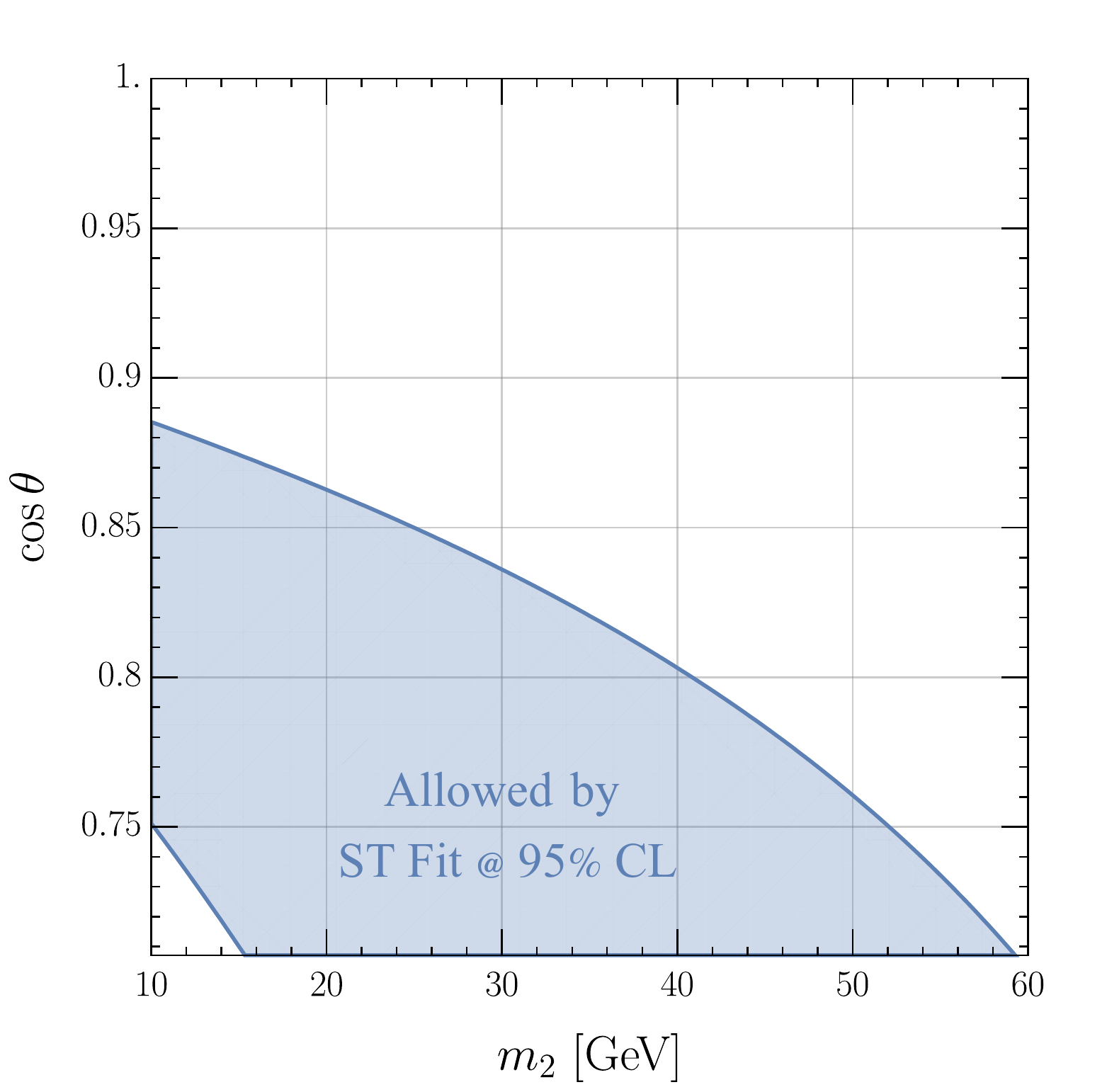}
	\end{center}
	\caption{The allowed parameter space at 95\% CL for xSM using the global fit from Ref.~\cite{deBlas:2022hdk} assuming $U=0$ and $M_W$ with conservative average. The vertical axis start from $\theta=\pi/4$, above which the mass eigenstate $h_1$ gets more contribution from the neutral part in the Higgs doublet thus corresponding to the SM-like Higgs.}\label{fig:ST}
\end{figure}

To illustrate the potential influence of such a new $m_W$ measurement on the EWPO constraint for xSM,  we follow the recent global fit result by the {\tt HEPfit} group~\cite{deBlas:2022hdk}, where they combine the measurements from LEP2, LHC and Tevatron including the most recent CDFII results to give a new ``world average'' of $W$ mass as:
\begin{eqnarray}
m_W = 80.4133 \pm  0.0080\ (0.015)\  {\rm GeV},
\end{eqnarray}
where the number in the parenthesis represents an inflated error for a conservative average.
From this updated value of $m_W$, authors in Ref.~\cite{deBlas:2022hdk} derive four sets of constraints on the $S,T$, and $U$ parameters assuming whether one uses the conservative average or not and whether one sets $U=0$ or not. As an example, we plot the allowed parameter space at 95\% confidence level (CL) in the $\cos\theta-m_{h_2}$ plane in the blue region in Fig.~\ref{fig:ST} assuming $U=0$ and $m_W$ taking the conservative average, which corresponds to the following bounds and correlations matrix on the $S$ and $T$ parameters:
\begin{eqnarray}
&&\Delta S=0.086\pm 0.077,\quad \Delta T=0.177\pm 0.070,\\
&&\rho = \begin{pmatrix}
1 & 0.89\\
0.89 & 1
\end{pmatrix}.
\end{eqnarray}
One can find that the allowed parameter space tends to have a very small $m_{h_2}$ such that the diHiggs decay channel is not available. This is expected, since the updated $W$ boson mass from CDFII indicates a relatively large violation of custodial symmetry, while xSM scalar potential does not explicitly break the custodial symmetry, as a consequence, the new physics contribution to the custodial symmetry breaking starts at one-loop receiving a loop factor suppression, to compensate such a suppression one needs a relatively small singlet-like Higgs mass. 
However, we need to emphasize here that the above global analysis heavily depends on the method used for combining the $W$ mass measurements from different experiments, given that the newly measured $m_W$ has a significant discrepancy compared with all the previous measurements, the ordinary averaging method assuming correlated Gaussian uncertainties may not be appropriate anymore, and the community has not reached a common agreement on the interpretation of the experimental results yet. 

\newpage
\vspace{0.5cm}
\phantomsection
\addcontentsline{toc}{section}{References}
\bibliography{xSM}

\providecommand{\href}[2]{#2}\begingroup\raggedright\begin{thebibliography}{10}

\bibitem{ATLAS:2012yve}
{\bfseries ATLAS} Collaboration, G.~Aad {\em et~al.}, ``{Observation of a new
  particle in the search for the Standard Model Higgs boson with the ATLAS
  detector at the LHC},''
  \href{http://dx.doi.org/10.1016/j.physletb.2012.08.020}{{\em Phys. Lett. B}
  {\bfseries 716} (2012) 1--29},
  \href{http://arxiv.org/abs/1207.7214}{{\ttfamily arXiv:1207.7214 [hep-ex]}}.

\bibitem{CMS:2012qbp}
{\bfseries CMS} Collaboration, S.~Chatrchyan {\em et~al.}, ``{Observation of a
  New Boson at a Mass of 125 GeV with the CMS Experiment at the LHC},''
  \href{http://dx.doi.org/10.1016/j.physletb.2012.08.021}{{\em Phys. Lett. B}
  {\bfseries 716} (2012) 30--61},
  \href{http://arxiv.org/abs/1207.7235}{{\ttfamily arXiv:1207.7235 [hep-ex]}}.

\bibitem{Planck:2013pxb}
{\bfseries Planck} Collaboration, P.~A.~R. Ade {\em et~al.}, ``{Planck 2013
  results. XVI. Cosmological parameters},''
  \href{http://dx.doi.org/10.1051/0004-6361/201321591}{{\em Astron. Astrophys.}
  {\bfseries 571} (2014) A16}, \href{http://arxiv.org/abs/1303.5076}{{\ttfamily
  arXiv:1303.5076 [astro-ph.CO]}}.

\bibitem{Kuzmin:1985mm}
V.~A. Kuzmin, V.~A. Rubakov, and M.~E. Shaposhnikov, ``{On the Anomalous
  Electroweak Baryon Number Nonconservation in the Early Universe},''
  \href{http://dx.doi.org/10.1016/0370-2693(85)91028-7}{{\em Phys. Lett. B}
  {\bfseries 155} (1985) 36}.

\bibitem{Shaposhnikov:1986jp}
M.~E. Shaposhnikov, ``{Possible Appearance of the Baryon Asymmetry of the
  Universe in an Electroweak Theory},'' {\em JETP Lett.} {\bfseries 44} (1986)
  465--468.

\bibitem{Shaposhnikov:1987tw}
M.~E. Shaposhnikov, ``{Baryon Asymmetry of the Universe in Standard Electroweak
  Theory},'' \href{http://dx.doi.org/10.1016/0550-3213(87)90127-1}{{\em Nucl.
  Phys. B} {\bfseries 287} (1987) 757--775}.

\bibitem{Cohen:1993nk}
A.~G. Cohen, D.~B. Kaplan, and A.~E. Nelson, ``{Progress in electroweak
  baryogenesis},''
  \href{http://dx.doi.org/10.1146/annurev.ns.43.120193.000331}{{\em Ann. Rev.
  Nucl. Part. Sci.} {\bfseries 43} (1993) 27--70},
  \href{http://arxiv.org/abs/hep-ph/9302210}{{\ttfamily arXiv:hep-ph/9302210}}.

\bibitem{Morrissey:2012db}
D.~E. Morrissey and M.~J. Ramsey-Musolf, ``{Electroweak baryogenesis},''
  \href{http://dx.doi.org/10.1088/1367-2630/14/12/125003}{{\em New J. Phys.}
  {\bfseries 14} (2012) 125003},
  \href{http://arxiv.org/abs/1206.2942}{{\ttfamily arXiv:1206.2942 [hep-ph]}}.

\bibitem{Sakharov:1967dj}
A.~D. Sakharov, ``{Violation of CP Invariance, C asymmetry, and baryon
  asymmetry of the universe},''
  \href{http://dx.doi.org/10.1070/PU1991v034n05ABEH002497}{{\em Pisma Zh. Eksp.
  Teor. Fiz.} {\bfseries 5} (1967) 32--35}.

\bibitem{Kajantie:1996qd}
K.~Kajantie, M.~Laine, K.~Rummukainen, and M.~E. Shaposhnikov, ``{A
  Nonperturbative analysis of the finite T phase transition in SU(2) x U(1)
  electroweak theory},''
  \href{http://dx.doi.org/10.1016/S0550-3213(97)00164-8}{{\em Nucl. Phys. B}
  {\bfseries 493} (1997) 413--438},
  \href{http://arxiv.org/abs/hep-lat/9612006}{{\ttfamily
  arXiv:hep-lat/9612006}}.

\bibitem{Klinkhamer:1984di}
F.~R. Klinkhamer and N.~S. Manton, ``{A Saddle Point Solution in the
  Weinberg-Salam Theory},''
  \href{http://dx.doi.org/10.1103/PhysRevD.30.2212}{{\em Phys. Rev. D}
  {\bfseries 30} (1984) 2212}.

\bibitem{Kajantie:1996mn}
K.~Kajantie, M.~Laine, K.~Rummukainen, and M.~E. Shaposhnikov, ``{Is there a
  hot electroweak phase transition at m(H) larger or equal to m(W)?},''
  \href{http://dx.doi.org/10.1103/PhysRevLett.77.2887}{{\em Phys. Rev. Lett.}
  {\bfseries 77} (1996) 2887--2890},
  \href{http://arxiv.org/abs/hep-ph/9605288}{{\ttfamily arXiv:hep-ph/9605288}}.

\bibitem{Ramsey-Musolf:2019lsf}
M.~J. Ramsey-Musolf, ``{The electroweak phase transition: a collider target},''
  \href{http://dx.doi.org/10.1007/JHEP09(2020)179}{{\em JHEP} {\bfseries 09}
  (2020) 179}, \href{http://arxiv.org/abs/1912.07189}{{\ttfamily
  arXiv:1912.07189 [hep-ph]}}.

\bibitem{OConnell:2006rsp}
D.~O'Connell, M.~J. Ramsey-Musolf, and M.~B. Wise, ``{Minimal Extension of the
  Standard Model Scalar Sector},''
  \href{http://dx.doi.org/10.1103/PhysRevD.75.037701}{{\em Phys. Rev. D}
  {\bfseries 75} (2007) 037701},
  \href{http://arxiv.org/abs/hep-ph/0611014}{{\ttfamily arXiv:hep-ph/0611014}}.

\bibitem{Profumo:2007wc}
S.~Profumo, M.~J. Ramsey-Musolf, and G.~Shaughnessy, ``{Singlet Higgs
  phenomenology and the electroweak phase transition},''
  \href{http://dx.doi.org/10.1088/1126-6708/2007/08/010}{{\em JHEP} {\bfseries
  08} (2007) 010}, \href{http://arxiv.org/abs/0705.2425}{{\ttfamily
  arXiv:0705.2425 [hep-ph]}}.

\bibitem{Kotwal:2016tex}
A.~V. Kotwal, M.~J. Ramsey-Musolf, J.~M. No, and P.~Winslow,
  ``{Singlet-catalyzed electroweak phase transitions in the 100 TeV
  frontier},'' \href{http://dx.doi.org/10.1103/PhysRevD.94.035022}{{\em Phys.
  Rev. D} {\bfseries 94} no.~3, (2016) 035022},
  \href{http://arxiv.org/abs/1605.06123}{{\ttfamily arXiv:1605.06123
  [hep-ph]}}.

\bibitem{Hashino:2016xoj}
K.~Hashino, M.~Kakizaki, S.~Kanemura, P.~Ko, and T.~Matsui, ``{Gravitational
  waves and Higgs boson couplings for exploring first order phase transition in
  the model with a singlet scalar field},''
  \href{http://dx.doi.org/10.1016/j.physletb.2016.12.052}{{\em Phys. Lett. B}
  {\bfseries 766} (2017) 49--54},
  \href{http://arxiv.org/abs/1609.00297}{{\ttfamily arXiv:1609.00297
  [hep-ph]}}.

\bibitem{ATLAS:2018rnh}
{\bfseries ATLAS} Collaboration, M.~Aaboud {\em et~al.}, ``{Search for pair
  production of Higgs bosons in the $b\bar{b}b\bar{b}$ final state using
  proton-proton collisions at $\sqrt{s} = 13$ TeV with the ATLAS detector},''
  \href{http://dx.doi.org/10.1007/JHEP01(2019)030}{{\em JHEP} {\bfseries 01}
  (2019) 030}, \href{http://arxiv.org/abs/1804.06174}{{\ttfamily
  arXiv:1804.06174 [hep-ex]}}.

\bibitem{CMS:2018qmt}
{\bfseries CMS} Collaboration, A.~M. Sirunyan {\em et~al.}, ``{Search for
  resonant pair production of Higgs bosons decaying to bottom quark-antiquark
  pairs in proton-proton collisions at 13 TeV},''
  \href{http://dx.doi.org/10.1007/JHEP08(2018)152}{{\em JHEP} {\bfseries 08}
  (2018) 152}, \href{http://arxiv.org/abs/1806.03548}{{\ttfamily
  arXiv:1806.03548 [hep-ex]}}.

\bibitem{ATLAS:2018fpd}
{\bfseries ATLAS} Collaboration, M.~Aaboud {\em et~al.}, ``{Search for Higgs
  boson pair production in the $b\bar{b}WW^{*}$ decay mode at $\sqrt{s}=13$ TeV
  with the ATLAS detector},''
  \href{http://dx.doi.org/10.1007/JHEP04(2019)092}{{\em JHEP} {\bfseries 04}
  (2019) 092}, \href{http://arxiv.org/abs/1811.04671}{{\ttfamily
  arXiv:1811.04671 [hep-ex]}}.

\bibitem{CMS:2017rpp}
{\bfseries CMS} Collaboration, A.~M. Sirunyan {\em et~al.}, ``{Search for
  resonant and nonresonant Higgs boson pair production in the $
  \mathrm{b}\overline{\mathrm{b}}\mathit{\ell \nu \ell \nu } $ final state in
  proton-proton collisions at $ \sqrt{s}=13 $ TeV},''
  \href{http://dx.doi.org/10.1007/JHEP01(2018)054}{{\em JHEP} {\bfseries 01}
  (2018) 054}, \href{http://arxiv.org/abs/1708.04188}{{\ttfamily
  arXiv:1708.04188 [hep-ex]}}.

\bibitem{ATLAS:2017tlw}
{\bfseries ATLAS} Collaboration, M.~Aaboud {\em et~al.}, ``{Search for heavy ZZ
  resonances in the $\ell ^+\ell ^-\ell ^+\ell ^-$ and $\ell ^+\ell ^-\nu
  \bar{\nu }$ final states using proton\textendash{}proton collisions at
  $\sqrt{s}= 13$ $\text {TeV}$ with the ATLAS detector},''
  \href{http://dx.doi.org/10.1140/epjc/s10052-018-5686-3}{{\em Eur. Phys. J. C}
  {\bfseries 78} no.~4, (2018) 293},
  \href{http://arxiv.org/abs/1712.06386}{{\ttfamily arXiv:1712.06386
  [hep-ex]}}.

\bibitem{ATLAS:2018ili}
{\bfseries ATLAS} Collaboration, M.~Aaboud {\em et~al.}, ``{Search for Higgs
  boson pair production in the $WW^{(*)}WW^{(*)}$ decay channel using ATLAS
  data recorded at $\sqrt{s}=13$ TeV},''
  \href{http://dx.doi.org/10.1007/JHEP05(2019)124}{{\em JHEP} {\bfseries 05}
  (2019) 124}, \href{http://arxiv.org/abs/1811.11028}{{\ttfamily
  arXiv:1811.11028 [hep-ex]}}.

\bibitem{ATLAS:2018uni}
{\bfseries ATLAS} Collaboration, M.~Aaboud {\em et~al.}, ``{Search for resonant
  and non-resonant Higgs boson pair production in the ${b\bar{b}\tau^+\tau^-}$
  decay channel in $pp$ collisions at $\sqrt{s}=13$ TeV with the ATLAS
  detector},'' \href{http://dx.doi.org/10.1103/PhysRevLett.121.191801}{{\em
  Phys. Rev. Lett.} {\bfseries 121} no.~19, (2018) 191801},
  \href{http://arxiv.org/abs/1808.00336}{{\ttfamily arXiv:1808.00336
  [hep-ex]}}. [Erratum: Phys.Rev.Lett. 122, 089901 (2019)].

\bibitem{CMS:2017hea}
{\bfseries CMS} Collaboration, A.~M. Sirunyan {\em et~al.}, ``{Search for Higgs
  boson pair production in events with two bottom quarks and two tau leptons in
  proton\textendash{}proton collisions at $\sqrt s$ =13TeV},''
  \href{http://dx.doi.org/10.1016/j.physletb.2018.01.001}{{\em Phys. Lett. B}
  {\bfseries 778} (2018) 101--127},
  \href{http://arxiv.org/abs/1707.02909}{{\ttfamily arXiv:1707.02909
  [hep-ex]}}.

\bibitem{ATLAS:2018dpp}
{\bfseries ATLAS} Collaboration, M.~Aaboud {\em et~al.}, ``{Search for Higgs
  boson pair production in the $\gamma\gamma b\bar{b}$ final state with 13 TeV
  $pp$ collision data collected by the ATLAS experiment},''
  \href{http://dx.doi.org/10.1007/JHEP11(2018)040}{{\em JHEP} {\bfseries 11}
  (2018) 040}, \href{http://arxiv.org/abs/1807.04873}{{\ttfamily
  arXiv:1807.04873 [hep-ex]}}.

\bibitem{CMS:2018tla}
{\bfseries CMS} Collaboration, A.~M. Sirunyan {\em et~al.}, ``{Search for Higgs
  boson pair production in the $\gamma\gamma\mathrm{b\overline{b}}$ final state
  in pp collisions at $\sqrt{s}=$ 13 TeV},''
  \href{http://dx.doi.org/10.1016/j.physletb.2018.10.056}{{\em Phys. Lett. B}
  {\bfseries 788} (2019) 7--36},
  \href{http://arxiv.org/abs/1806.00408}{{\ttfamily arXiv:1806.00408
  [hep-ex]}}.

\bibitem{ATLAS:2020fry}
{\bfseries ATLAS} Collaboration, G.~Aad {\em et~al.}, ``{Search for heavy
  diboson resonances in semileptonic final states in pp collisions at
  $\sqrt{s}=13$ TeV with the ATLAS detector},''
  \href{http://dx.doi.org/10.1140/epjc/s10052-020-08554-y}{{\em Eur. Phys. J.
  C} {\bfseries 80} no.~12, (2020) 1165},
  \href{http://arxiv.org/abs/2004.14636}{{\ttfamily arXiv:2004.14636
  [hep-ex]}}.

\bibitem{ATLAS:2017jag}
{\bfseries ATLAS} Collaboration, M.~Aaboud {\em et~al.}, ``{Search for $WW/WZ$
  resonance production in $\ell \nu qq$ final states in $pp$ collisions at
  $\sqrt{s} =$ 13 TeV with the ATLAS detector},''
  \href{http://dx.doi.org/10.1007/JHEP03(2018)042}{{\em JHEP} {\bfseries 03}
  (2018) 042}, \href{http://arxiv.org/abs/1710.07235}{{\ttfamily
  arXiv:1710.07235 [hep-ex]}}.

\bibitem{ATLAS:2016hal}
{\bfseries ATLAS} Collaboration, M.~Aaboud {\em et~al.}, ``{Searches for heavy
  diboson resonances in $pp$ collisions at $\sqrt{s}=13$ TeV with the ATLAS
  detector},'' \href{http://dx.doi.org/10.1007/JHEP09(2016)173}{{\em JHEP}
  {\bfseries 09} (2016) 173}, \href{http://arxiv.org/abs/1606.04833}{{\ttfamily
  arXiv:1606.04833 [hep-ex]}}.

\bibitem{ATLAS:2017zuf}
{\bfseries ATLAS} Collaboration, M.~Aaboud {\em et~al.}, ``{Search for diboson
  resonances with boson-tagged jets in $pp$ collisions at $\sqrt{s}=13$ TeV
  with the ATLAS detector},''
  \href{http://dx.doi.org/10.1016/j.physletb.2017.12.011}{{\em Phys. Lett. B}
  {\bfseries 777} (2018) 91--113},
  \href{http://arxiv.org/abs/1708.04445}{{\ttfamily arXiv:1708.04445
  [hep-ex]}}.

\bibitem{ATLAS:2018sbw}
{\bfseries ATLAS} Collaboration, M.~Aaboud {\em et~al.}, ``{Combination of
  searches for heavy resonances decaying into bosonic and leptonic final states
  using 36 fb$^{-1}$ of proton-proton collision data at $\sqrt{s} = 13$ TeV
  with the ATLAS detector},''
  \href{http://dx.doi.org/10.1103/PhysRevD.98.052008}{{\em Phys. Rev. D}
  {\bfseries 98} no.~5, (2018) 052008},
  \href{http://arxiv.org/abs/1808.02380}{{\ttfamily arXiv:1808.02380
  [hep-ex]}}.

\bibitem{ATLAS:2017uhp}
{\bfseries ATLAS} Collaboration, M.~Aaboud {\em et~al.}, ``{Search for heavy
  resonances decaying into $WW$ in the $e\nu\mu\nu$ final state in $pp$
  collisions at $\sqrt{s}=13$ TeV with the ATLAS detector},''
  \href{http://dx.doi.org/10.1140/epjc/s10052-017-5491-4}{{\em Eur. Phys. J. C}
  {\bfseries 78} no.~1, (2018) 24},
  \href{http://arxiv.org/abs/1710.01123}{{\ttfamily arXiv:1710.01123
  [hep-ex]}}.

\bibitem{ATLAS:2020zms}
{\bfseries ATLAS} Collaboration, G.~Aad {\em et~al.}, ``{Search for heavy Higgs
  bosons decaying into two tau leptons with the ATLAS detector using $pp$
  collisions at $\sqrt{s}=13$ TeV},''
  \href{http://dx.doi.org/10.1103/PhysRevLett.125.051801}{{\em Phys. Rev.
  Lett.} {\bfseries 125} no.~5, (2020) 051801},
  \href{http://arxiv.org/abs/2002.12223}{{\ttfamily arXiv:2002.12223
  [hep-ex]}}.

\bibitem{Huang:2017jws}
T.~Huang, J.~M. No, L.~Perni\'e, M.~Ramsey-Musolf, A.~Safonov, M.~Spannowsky,
  and P.~Winslow, ``{Resonant di-Higgs boson production in the $b{\bar b}WW$
  channel: Probing the electroweak phase transition at the LHC},''
  \href{http://dx.doi.org/10.1103/PhysRevD.96.035007}{{\em Phys. Rev. D}
  {\bfseries 96} no.~3, (2017) 035007},
  \href{http://arxiv.org/abs/1701.04442}{{\ttfamily arXiv:1701.04442
  [hep-ph]}}.

\bibitem{Espinosa:2011ax}
J.~R. Espinosa, T.~Konstandin, and F.~Riva, ``{Strong Electroweak Phase
  Transitions in the Standard Model with a Singlet},''
  \href{http://dx.doi.org/10.1016/j.nuclphysb.2011.09.010}{{\em Nucl. Phys. B}
  {\bfseries 854} (2012) 592--630},
  \href{http://arxiv.org/abs/1107.5441}{{\ttfamily arXiv:1107.5441 [hep-ph]}}.

\bibitem{Patel:2011th}
H.~H. Patel and M.~J. Ramsey-Musolf, ``{Baryon Washout, Electroweak Phase
  Transition, and Perturbation Theory},''
  \href{http://dx.doi.org/10.1007/JHEP07(2011)029}{{\em JHEP} {\bfseries 07}
  (2011) 029}, \href{http://arxiv.org/abs/1101.4665}{{\ttfamily arXiv:1101.4665
  [hep-ph]}}.

\bibitem{Laine:1994zq}
M.~Laine, ``{Gauge dependence of the high temperature two loop effective
  potential for the Higgs field},''
  \href{http://dx.doi.org/10.1103/PhysRevD.51.4525}{{\em Phys. Rev. D}
  {\bfseries 51} (1995) 4525--4532},
  \href{http://arxiv.org/abs/hep-ph/9411252}{{\ttfamily arXiv:hep-ph/9411252}}.

\bibitem{Ekstedt:2018ftj}
A.~Ekstedt and J.~L\"ofgren, ``{On the relationship between gauge dependence
  and IR divergences in the $\hbar$-expansion of the effective potential},''
  \href{http://dx.doi.org/10.1007/JHEP01(2019)226}{{\em JHEP} {\bfseries 01}
  (2019) 226}, \href{http://arxiv.org/abs/1810.01416}{{\ttfamily
  arXiv:1810.01416 [hep-ph]}}.

\bibitem{Garny:2012cg}
M.~Garny and T.~Konstandin, ``{On the gauge dependence of vacuum transitions at
  finite temperature},'' \href{http://dx.doi.org/10.1007/JHEP07(2012)189}{{\em
  JHEP} {\bfseries 07} (2012) 189},
  \href{http://arxiv.org/abs/1205.3392}{{\ttfamily arXiv:1205.3392 [hep-ph]}}.

\bibitem{Carena:2019une}
M.~Carena, Z.~Liu, and Y.~Wang, ``{Electroweak phase transition with
  spontaneous Z$_{2}$-breaking},''
  \href{http://dx.doi.org/10.1007/JHEP08(2020)107}{{\em JHEP} {\bfseries 08}
  (2020) 107}, \href{http://arxiv.org/abs/1911.10206}{{\ttfamily
  arXiv:1911.10206 [hep-ph]}}.

\bibitem{Kozaczuk:2019pet}
J.~Kozaczuk, M.~J. Ramsey-Musolf, and J.~Shelton, ``{Exotic Higgs boson decays
  and the electroweak phase transition},''
  \href{http://dx.doi.org/10.1103/PhysRevD.101.115035}{{\em Phys. Rev. D}
  {\bfseries 101} no.~11, (2020) 115035},
  \href{http://arxiv.org/abs/1911.10210}{{\ttfamily arXiv:1911.10210
  [hep-ph]}}.

\bibitem{Alves:2018jsw}
A.~Alves, T.~Ghosh, H.-K. Guo, K.~Sinha, and D.~Vagie, ``{Collider and
  Gravitational Wave Complementarity in Exploring the Singlet Extension of the
  Standard Model},'' \href{http://dx.doi.org/10.1007/JHEP04(2019)052}{{\em
  JHEP} {\bfseries 04} (2019) 052},
  \href{http://arxiv.org/abs/1812.09333}{{\ttfamily arXiv:1812.09333
  [hep-ph]}}.

\bibitem{Kurup:2017dzf}
G.~Kurup and M.~Perelstein, ``{Dynamics of Electroweak Phase Transition In
  Singlet-Scalar Extension of the Standard Model},''
  \href{http://dx.doi.org/10.1103/PhysRevD.96.015036}{{\em Phys. Rev. D}
  {\bfseries 96} no.~1, (2017) 015036},
  \href{http://arxiv.org/abs/1704.03381}{{\ttfamily arXiv:1704.03381
  [hep-ph]}}.

\bibitem{Chen:2017qcz}
C.-Y. Chen, J.~Kozaczuk, and I.~M. Lewis, ``{Non-resonant Collider Signatures
  of a Singlet-Driven Electroweak Phase Transition},''
  \href{http://dx.doi.org/10.1007/JHEP08(2017)096}{{\em JHEP} {\bfseries 08}
  (2017) 096}, \href{http://arxiv.org/abs/1704.05844}{{\ttfamily
  arXiv:1704.05844 [hep-ph]}}.

\bibitem{Chen:2014ask}
C.-Y. Chen, S.~Dawson, and I.~M. Lewis, ``{Exploring resonant di-Higgs boson
  production in the Higgs singlet model},''
  \href{http://dx.doi.org/10.1103/PhysRevD.91.035015}{{\em Phys. Rev. D}
  {\bfseries 91} no.~3, (2015) 035015},
  \href{http://arxiv.org/abs/1410.5488}{{\ttfamily arXiv:1410.5488 [hep-ph]}}.

\bibitem{Katz:2014bha}
A.~Katz and M.~Perelstein, ``{Higgs Couplings and Electroweak Phase
  Transition},'' \href{http://dx.doi.org/10.1007/JHEP07(2014)108}{{\em JHEP}
  {\bfseries 07} (2014) 108}, \href{http://arxiv.org/abs/1401.1827}{{\ttfamily
  arXiv:1401.1827 [hep-ph]}}.

\bibitem{Damgaard:2015con}
P.~H. Damgaard, A.~Haarr, D.~O'Connell, and A.~Tranberg, ``{Effective Field
  Theory and Electroweak Baryogenesis in the Singlet-Extended Standard
  Model},'' \href{http://dx.doi.org/10.1007/JHEP02(2016)107}{{\em JHEP}
  {\bfseries 02} (2016) 107}, \href{http://arxiv.org/abs/1512.01963}{{\ttfamily
  arXiv:1512.01963 [hep-ph]}}.

\bibitem{Huang:2015tdv}
P.~Huang, A.~Joglekar, B.~Li, and C.~E.~M. Wagner, ``{Probing the Electroweak
  Phase Transition at the LHC},''
  \href{http://dx.doi.org/10.1103/PhysRevD.93.055049}{{\em Phys. Rev. D}
  {\bfseries 93} no.~5, (2016) 055049},
  \href{http://arxiv.org/abs/1512.00068}{{\ttfamily arXiv:1512.00068
  [hep-ph]}}.

\bibitem{Curtin:2014jma}
D.~Curtin, P.~Meade, and C.-T. Yu, ``{Testing Electroweak Baryogenesis with
  Future Colliders},'' \href{http://dx.doi.org/10.1007/JHEP11(2014)127}{{\em
  JHEP} {\bfseries 11} (2014) 127},
  \href{http://arxiv.org/abs/1409.0005}{{\ttfamily arXiv:1409.0005 [hep-ph]}}.

\bibitem{No:2013wsa}
J.~M. No and M.~Ramsey-Musolf, ``{Probing the Higgs Portal at the LHC Through
  Resonant di-Higgs Production},''
  \href{http://dx.doi.org/10.1103/PhysRevD.89.095031}{{\em Phys. Rev. D}
  {\bfseries 89} no.~9, (2014) 095031},
  \href{http://arxiv.org/abs/1310.6035}{{\ttfamily arXiv:1310.6035 [hep-ph]}}.

\bibitem{Huang:2012wn}
W.~Huang, J.~Shu, and Y.~Zhang, ``{On the Higgs Fit and Electroweak Phase
  Transition},'' \href{http://dx.doi.org/10.1007/JHEP03(2013)164}{{\em JHEP}
  {\bfseries 03} (2013) 164}, \href{http://arxiv.org/abs/1210.0906}{{\ttfamily
  arXiv:1210.0906 [hep-ph]}}.

\bibitem{Damgaard:2013kva}
P.~H. Damgaard, D.~O'Connell, T.~C. Petersen, and A.~Tranberg, ``{Constraints
  on New Physics from Baryogenesis and Large Hadron Collider Data},''
  \href{http://dx.doi.org/10.1103/PhysRevLett.111.221804}{{\em Phys. Rev.
  Lett.} {\bfseries 111} no.~22, (2013) 221804},
  \href{http://arxiv.org/abs/1305.4362}{{\ttfamily arXiv:1305.4362 [hep-ph]}}.

\bibitem{Ashoorioon:2009nf}
A.~Ashoorioon and T.~Konstandin, ``{Strong electroweak phase transitions
  without collider traces},''
  \href{http://dx.doi.org/10.1088/1126-6708/2009/07/086}{{\em JHEP} {\bfseries
  07} (2009) 086}, \href{http://arxiv.org/abs/0904.0353}{{\ttfamily
  arXiv:0904.0353 [hep-ph]}}.

\bibitem{Blasi:2022woz}
S.~Blasi and A.~Mariotti, ``{Domain Walls Seeding the Electroweak Phase
  Transition},'' \href{http://dx.doi.org/10.1103/PhysRevLett.129.261303}{{\em
  Phys. Rev. Lett.} {\bfseries 129} no.~26, (2022) 261303},
  \href{http://arxiv.org/abs/2203.16450}{{\ttfamily arXiv:2203.16450
  [hep-ph]}}.

\bibitem{Ghorbani:2020xqv}
P.~Ghorbani, ``{Vacuum structure and electroweak phase transition in singlet
  scalar dark matter},''
  \href{http://dx.doi.org/10.1016/j.dark.2021.100861}{{\em Phys. Dark Univ.}
  {\bfseries 33} (2021) 100861},
  \href{http://arxiv.org/abs/2010.15708}{{\ttfamily arXiv:2010.15708
  [hep-ph]}}.

\bibitem{Ghorbani:2018yfr}
K.~Ghorbani and P.~H. Ghorbani, ``{Strongly First-Order Phase Transition in
  Real Singlet Scalar Dark Matter Model},''
  \href{http://dx.doi.org/10.1088/1361-6471/ab4823}{{\em J. Phys. G} {\bfseries
  47} no.~1, (2020) 015201}, \href{http://arxiv.org/abs/1804.05798}{{\ttfamily
  arXiv:1804.05798 [hep-ph]}}.

\bibitem{Noble:2007kk}
A.~Noble and M.~Perelstein, ``{Higgs self-coupling as a probe of electroweak
  phase transition},'' \href{http://dx.doi.org/10.1103/PhysRevD.78.063518}{{\em
  Phys. Rev. D} {\bfseries 78} (2008) 063518},
  \href{http://arxiv.org/abs/0711.3018}{{\ttfamily arXiv:0711.3018 [hep-ph]}}.

\bibitem{Espinosa:2007qk}
J.~R. Espinosa and M.~Quiros, ``{Novel Effects in Electroweak Breaking from a
  Hidden Sector},'' \href{http://dx.doi.org/10.1103/PhysRevD.76.076004}{{\em
  Phys. Rev. D} {\bfseries 76} (2007) 076004},
  \href{http://arxiv.org/abs/hep-ph/0701145}{{\ttfamily arXiv:hep-ph/0701145}}.

\bibitem{Cline:2012hg}
J.~M. Cline and K.~Kainulainen, ``{Electroweak baryogenesis and dark matter
  from a singlet Higgs},''
  \href{http://dx.doi.org/10.1088/1475-7516/2013/01/012}{{\em JCAP} {\bfseries
  01} (2013) 012}, \href{http://arxiv.org/abs/1210.4196}{{\ttfamily
  arXiv:1210.4196 [hep-ph]}}.

\bibitem{Profumo:2014opa}
S.~Profumo, M.~J. Ramsey-Musolf, C.~L. Wainwright, and P.~Winslow,
  ``{Singlet-catalyzed electroweak phase transitions and precision Higgs boson
  studies},'' \href{http://dx.doi.org/10.1103/PhysRevD.91.035018}{{\em Phys.
  Rev. D} {\bfseries 91} no.~3, (2015) 035018},
  \href{http://arxiv.org/abs/1407.5342}{{\ttfamily arXiv:1407.5342 [hep-ph]}}.

\bibitem{ParticleDataGroup:2022pth}
{\bfseries Particle Data Group} Collaboration, R.~L. Workman {\em et~al.},
  ``{Review of Particle Physics},''
  \href{http://dx.doi.org/10.1093/ptep/ptac097}{{\em PTEP} {\bfseries 2022}
  (2022) 083C01}.

\bibitem{Wainwright:2011kj}
C.~L. Wainwright, ``{CosmoTransitions: Computing Cosmological Phase Transition
  Temperatures and Bubble Profiles with Multiple Fields},''
  \href{http://dx.doi.org/10.1016/j.cpc.2012.04.004}{{\em Comput. Phys.
  Commun.} {\bfseries 183} (2012) 2006--2013},
  \href{http://arxiv.org/abs/1109.4189}{{\ttfamily arXiv:1109.4189 [hep-ph]}}.

\bibitem{Lerner:2009xg}
R.~N. Lerner and J.~McDonald, ``{Gauge singlet scalar as inflaton and thermal
  relic dark matter},''
  \href{http://dx.doi.org/10.1103/PhysRevD.80.123507}{{\em Phys. Rev. D}
  {\bfseries 80} (2009) 123507},
  \href{http://arxiv.org/abs/0909.0520}{{\ttfamily arXiv:0909.0520 [hep-ph]}}.

\bibitem{Li:2019tfd}
H.-L. Li, M.~Ramsey-Musolf, and S.~Willocq, ``{Probing a scalar
  singlet-catalyzed electroweak phase transition with resonant di-Higgs boson
  production in the $4b$ channel},''
  \href{http://dx.doi.org/10.1103/PhysRevD.100.075035}{{\em Phys. Rev. D}
  {\bfseries 100} no.~7, (2019) 075035},
  \href{http://arxiv.org/abs/1906.05289}{{\ttfamily arXiv:1906.05289
  [hep-ph]}}.

\bibitem{ATLAS:2017rzl}
{\bfseries ATLAS} Collaboration, M.~Aaboud {\em et~al.}, ``{Measurement of the
  $W$-boson mass in pp collisions at $\sqrt{s}=7$ TeV with the ATLAS
  detector},'' \href{http://dx.doi.org/10.1140/epjc/s10052-017-5475-4}{{\em
  Eur. Phys. J. C} {\bfseries 78} no.~2, (2018) 110},
  \href{http://arxiv.org/abs/1701.07240}{{\ttfamily arXiv:1701.07240
  [hep-ex]}}. [Erratum: Eur.Phys.J.C 78, 898 (2018)].

\bibitem{ATLAS:2022vkf}
{\bfseries ATLAS} Collaboration, ``{A detailed map of Higgs boson interactions
  by the ATLAS experiment ten years after the discovery},''
  \href{http://dx.doi.org/10.1038/s41586-022-04893-w}{{\em Nature} {\bfseries
  607} no.~7917, (2022) 52--59},
  \href{http://arxiv.org/abs/2207.00092}{{\ttfamily arXiv:2207.00092
  [hep-ex]}}. [Erratum: Nature 612, E24 (2022)].

\bibitem{CMS:2020gsy}
{\bfseries CMS} Collaboration, ``{Combined Higgs boson production and decay
  measurements with up to 137 fb$^{-1}$ of proton-proton collision data at
  $\sqrt s$ = 13 TeV},''.

\bibitem{Cepeda:2019klc}
M.~Cepeda {\em et~al.}, ``{Report from Working Group 2}: {Higgs Physics at the
  HL-LHC and HE-LHC},''
  \href{http://dx.doi.org/10.23731/CYRM-2019-007.221}{{\em CERN Yellow Rep.
  Monogr.} {\bfseries 7} (2019) 221--584},
  \href{http://arxiv.org/abs/1902.00134}{{\ttfamily arXiv:1902.00134
  [hep-ph]}}.

\bibitem{pyhf}
L.~Heinrich, M.~Feickert, and G.~Stark, ``{pyhf: v0.7.3}.''
\newblock \url{https://doi.org/10.5281/zenodo.1169739}.
  https://github.com/scikit-hep/pyhf/releases/tag/v0.7.3.

\bibitem{pyhf_joss}
L.~Heinrich, M.~Feickert, G.~Stark, and K.~Cranmer, ``pyhf: pure-python
  implementation of histfactory statistical models,''
  \href{http://dx.doi.org/10.21105/joss.02823}{{\em Journal of Open Source
  Software} {\bfseries 6} no.~58, (2021) 2823}.
  \url{https://doi.org/10.21105/joss.02823}.

\bibitem{Sartore:2020gou}
L.~Sartore and I.~Schienbein, ``{PyR@TE 3},''
  \href{http://dx.doi.org/10.1016/j.cpc.2020.107819}{{\em Comput. Phys.
  Commun.} {\bfseries 261} (2021) 107819},
  \href{http://arxiv.org/abs/2007.12700}{{\ttfamily arXiv:2007.12700
  [hep-ph]}}.

\bibitem{Schienbein:2018fsw}
I.~Schienbein, F.~Staub, T.~Steudtner, and K.~Svirina, ``{Revisiting RGEs for
  general gauge theories},''
  \href{http://dx.doi.org/10.1016/j.nuclphysb.2018.12.001}{{\em Nucl. Phys. B}
  {\bfseries 939} (2019) 1--48},
  \href{http://arxiv.org/abs/1809.06797}{{\ttfamily arXiv:1809.06797
  [hep-ph]}}. [Erratum: Nucl.Phys.B 966, 115339 (2021)].

\bibitem{Luo:2002ti}
M.-x. Luo, H.-w. Wang, and Y.~Xiao, ``{Two loop renormalization group equations
  in general gauge field theories},''
  \href{http://dx.doi.org/10.1103/PhysRevD.67.065019}{{\em Phys. Rev. D}
  {\bfseries 67} (2003) 065019},
  \href{http://arxiv.org/abs/hep-ph/0211440}{{\ttfamily arXiv:hep-ph/0211440}}.

\bibitem{Luo:2002ey}
M.-x. Luo and Y.~Xiao, ``{Two loop renormalization group equations in the
  standard model},''
  \href{http://dx.doi.org/10.1103/PhysRevLett.90.011601}{{\em Phys. Rev. Lett.}
  {\bfseries 90} (2003) 011601},
  \href{http://arxiv.org/abs/hep-ph/0207271}{{\ttfamily arXiv:hep-ph/0207271}}.

\bibitem{Gonderinger:2009jp}
M.~Gonderinger, Y.~Li, H.~Patel, and M.~J. Ramsey-Musolf, ``{Vacuum Stability,
  Perturbativity, and Scalar Singlet Dark Matter},''
  \href{http://dx.doi.org/10.1007/JHEP01(2010)053}{{\em JHEP} {\bfseries 01}
  (2010) 053}, \href{http://arxiv.org/abs/0910.3167}{{\ttfamily arXiv:0910.3167
  [hep-ph]}}.

\bibitem{Kniehl:2016enc}
B.~A. Kniehl, A.~F. Pikelner, and O.~L. Veretin, ``{mr: a C++ library for the
  matching and running of the Standard Model parameters},''
  \href{http://dx.doi.org/10.1016/j.cpc.2016.04.017}{{\em Comput. Phys.
  Commun.} {\bfseries 206} (2016) 84--96},
  \href{http://arxiv.org/abs/1601.08143}{{\ttfamily arXiv:1601.08143
  [hep-ph]}}.

\bibitem{Alwall:2011uj}
J.~Alwall, M.~Herquet, F.~Maltoni, O.~Mattelaer, and T.~Stelzer, ``{MadGraph 5
  : Going Beyond},'' \href{http://dx.doi.org/10.1007/JHEP06(2011)128}{{\em
  JHEP} {\bfseries 06} (2011) 128},
  \href{http://arxiv.org/abs/1106.0522}{{\ttfamily arXiv:1106.0522 [hep-ph]}}.

\bibitem{Sjostrand:2007gs}
T.~Sjostrand, S.~Mrenna, and P.~Z. Skands, ``{A Brief Introduction to PYTHIA
  8.1},'' \href{http://dx.doi.org/10.1016/j.cpc.2008.01.036}{{\em Comput. Phys.
  Commun.} {\bfseries 178} (2008) 852--867},
  \href{http://arxiv.org/abs/0710.3820}{{\ttfamily arXiv:0710.3820 [hep-ph]}}.

\bibitem{deFavereau:2013fsa}
{\bfseries DELPHES 3} Collaboration, J.~de~Favereau, C.~Delaere, P.~Demin,
  A.~Giammanco, V.~Lema\^\i{}tre, A.~Mertens, and M.~Selvaggi, ``{DELPHES 3, A
  modular framework for fast simulation of a generic collider experiment},''
  \href{http://dx.doi.org/10.1007/JHEP02(2014)057}{{\em JHEP} {\bfseries 02}
  (2014) 057}, \href{http://arxiv.org/abs/1307.6346}{{\ttfamily arXiv:1307.6346
  [hep-ex]}}.

\bibitem{Cacciari:2008gp}
M.~Cacciari, G.~P. Salam, and G.~Soyez, ``{The anti-$k_t$ jet clustering
  algorithm},'' \href{http://dx.doi.org/10.1088/1126-6708/2008/04/063}{{\em
  JHEP} {\bfseries 04} (2008) 063},
  \href{http://arxiv.org/abs/0802.1189}{{\ttfamily arXiv:0802.1189 [hep-ph]}}.

\bibitem{CMS:2012feb}
{\bfseries CMS} Collaboration, S.~Chatrchyan {\em et~al.}, ``{Identification of
  b-Quark Jets with the CMS Experiment},''
  \href{http://dx.doi.org/10.1088/1748-0221/8/04/P04013}{{\em JINST} {\bfseries
  8} (2013) P04013}, \href{http://arxiv.org/abs/1211.4462}{{\ttfamily
  arXiv:1211.4462 [hep-ex]}}.

\bibitem{Peskin:1990zt}
M.~E. Peskin and T.~Takeuchi, ``{A New constraint on a strongly interacting
  Higgs sector},'' \href{http://dx.doi.org/10.1103/PhysRevLett.65.964}{{\em
  Phys. Rev. Lett.} {\bfseries 65} (1990) 964--967}.

\bibitem{Peskin:1991sw}
M.~E. Peskin and T.~Takeuchi, ``{Estimation of oblique electroweak
  corrections},'' \href{http://dx.doi.org/10.1103/PhysRevD.46.381}{{\em Phys.
  Rev. D} {\bfseries 46} (1992) 381--409}.

\bibitem{Baak:2014ora}
{\bfseries Gfitter Group} Collaboration, M.~Baak, J.~C\'uth, J.~Haller,
  A.~Hoecker, R.~Kogler, K.~M\"onig, M.~Schott, and J.~Stelzer, ``{The global
  electroweak fit at NNLO and prospects for the LHC and ILC},''
  \href{http://dx.doi.org/10.1140/epjc/s10052-014-3046-5}{{\em Eur. Phys. J. C}
  {\bfseries 74} (2014) 3046}, \href{http://arxiv.org/abs/1407.3792}{{\ttfamily
  arXiv:1407.3792 [hep-ph]}}.

\bibitem{CDF:2022hxs}
{\bfseries CDF} Collaboration, T.~Aaltonen {\em et~al.}, ``{High-precision
  measurement of the W boson mass with the CDF II detector},''
  \href{http://dx.doi.org/10.1126/science.abk1781}{{\em Science} {\bfseries
  376} no.~6589, (2022) 170--176}.

\bibitem{deBlas:2022hdk}
J.~de~Blas, M.~Pierini, L.~Reina, and L.~Silvestrini, ``{Impact of the recent
  measurements of the top-quark and W-boson masses on electroweak precision
  fits},'' \href{http://arxiv.org/abs/2204.04204}{{\ttfamily arXiv:2204.04204
  [hep-ph]}}.

\end{thebibliography}\endgroup
\bibliographystyle{utphys}	

\end{document}